\documentclass[manuscript]{acmart}
\usepackage{graphicx}
\usepackage{adjustbox}
\usepackage{subfigure}

\usepackage{todonotes} 
\usepackage{multirow}
\AtBeginDocument{%
  \providecommand\BibTeX{{%
    \normalfont B\kern-0.5em{\scshape i\kern-0.25em b}\kern-0.8em\TeX}}}

\setcopyright{acmcopyright}
\copyrightyear{2023}
\acmYear{2023}
\acmDOI{0000000000}

\acmConference{}
\acmPrice{}
\acmISBN{}

\begin{document}

\title{Prompting Datasets: Data Discovery with Conversational Agents }

\author{Johanna Walker}

\email{johanna.walker@kcl.ac.uk}
\orcid{0000-0002-5498-8670}

\authornotemark[1]
\email{johanna.walker@kcl.ac.uk}
\affiliation{%
  \institution{King's College London}
  \streetaddress{Bush House, Strand}
  \city{London}
  \state{}
  \country{United Kingdom}
  \postcode{WC2B 4BG}
}

\author{Elisavet Koutsiana}

\email{elisavet.koutsiana@kcl.ac.uk}
\orcid{0000-0001-6544-0435}

\authornotemark[1]
\email{elisavet.koutsiana@kcl.ac.uk}
\affiliation{%
  \institution{King's College London}
  \streetaddress{Bush House, Strand}
  \city{London}
  \state{}
  \country{United Kingdom}
  \postcode{WC2B 4BG}
}

\author{Joe Massey}

\email{Joe.Massey@theodi.org}
\orcid{0009-0004-2880-5596}

\authornotemark[2]
\email{Joe.Massey@theodi.org}
\affiliation{%
  \institution{Open Data Institute}
  \streetaddress{4th Floor, Kings Place, 90 York Way}
  \city{London}
  \state{}
  \country{United Kingdom}
  \postcode{N1 9AG}
}

\author{Gefion Theurmer}

\email{gefion.thuermer@kcl.ac.uk}
\orcid{0000-0001-7345-0000}

\authornotemark[1]
\email{gefion.thuermer@kcl.ac.uk}
\affiliation{%
  \institution{King's College London}
  \streetaddress{Bush House, Strand}
  \city{London}
  \state{}
  \country{United Kingdom}
  \postcode{WC2B 4BG}
}

\author{Elena Simperl}

\email{elena.simperl@kcl.ac.uk}
\orcid{0000-0003-1722-947X}

\authornotemark[1]
\email{elena.simperl@kcl.ac.uk}
\affiliation{%
  \institution{King's College London}
  \streetaddress{Bush House, Strand}
  \city{London}
  \state{}
  \country{United Kingdom}
  \postcode{WC2B 4BG}
}

\renewcommand{\shortauthors}{Walker, J. et al.}

\begin{abstract}
  Can large language models assist in data discovery? Data discovery predominantly happens via search on a data portal or the web, followed by assessment of the dataset to ensure it is fit for the intended purpose. The ability of conversational generative AI (CGAI) to support recommendations with reasoning implies it can suggest datasets to users, explain why it has done so, and provide information akin to documentation regarding the dataset in order to support a use decision. We hold 3 workshops with  data users and find that, despite limitations around web capabilities, CGAIs are able to suggest relevant datasets and provide many of the required sensemaking activities, as well as support dataset analysis and manipulation. However, CGAIs may also suggest fictional datasets, and perform inaccurate analysis.  We identify emerging practices in data discovery and present a model of these to inform future research directions and data prompt design. 

\end{abstract}




\maketitle

\section{Introduction}
Dataset discovery, or identifying secondary data useful for research, business, or individual purposes, is a vital activity in contemporary life \cite{borner2018skill,jia2022truth}. This is increasingly true for not only data experts, but also for more casual users, who typically do not require large scale, well promoted datasets \cite{borner2018skill}. Dataset discovery is composed of two parts: searching for data, and assessing the data to ensure it is fit for the intended purpose. This second part is often characterized as sense-making \cite{chapman2020dataset}. The predominant methods of discovery currently are searching on a data portal (e.g. data.europa.eu) or via a search engine, including specialist search such as Google Dataset Search \cite{ibanez2022comparison}. 

Making data useable requires tools that support individuals’ ``data journey'' \cite{koesten2021ux,kacprzak2019characterising}. This includes specifying a goal and finding the data they need; exploring new sources of data; assessing the data's relevance and credibility, and deriving insights. Additionally, to support non-expert users, such tools need  to function effectively in low- to no-code interactions that are suitable to populations with diverse economic, cultural and educational backgrounds \cite{hayes2023human,schneider2023investigating}.   

Current data journeys involve creating ad hoc workflow processes on the fly over a number of tools, many of which are not easy for non-experts to use \cite{borner2018skill}. A typical set of search and sense-making activities might look as shown in Figure \ref{fig:five_figures}:

\begin{figure}[h]
    \centering
    \subfigure[]{\includegraphics[width=0.45\textwidth]{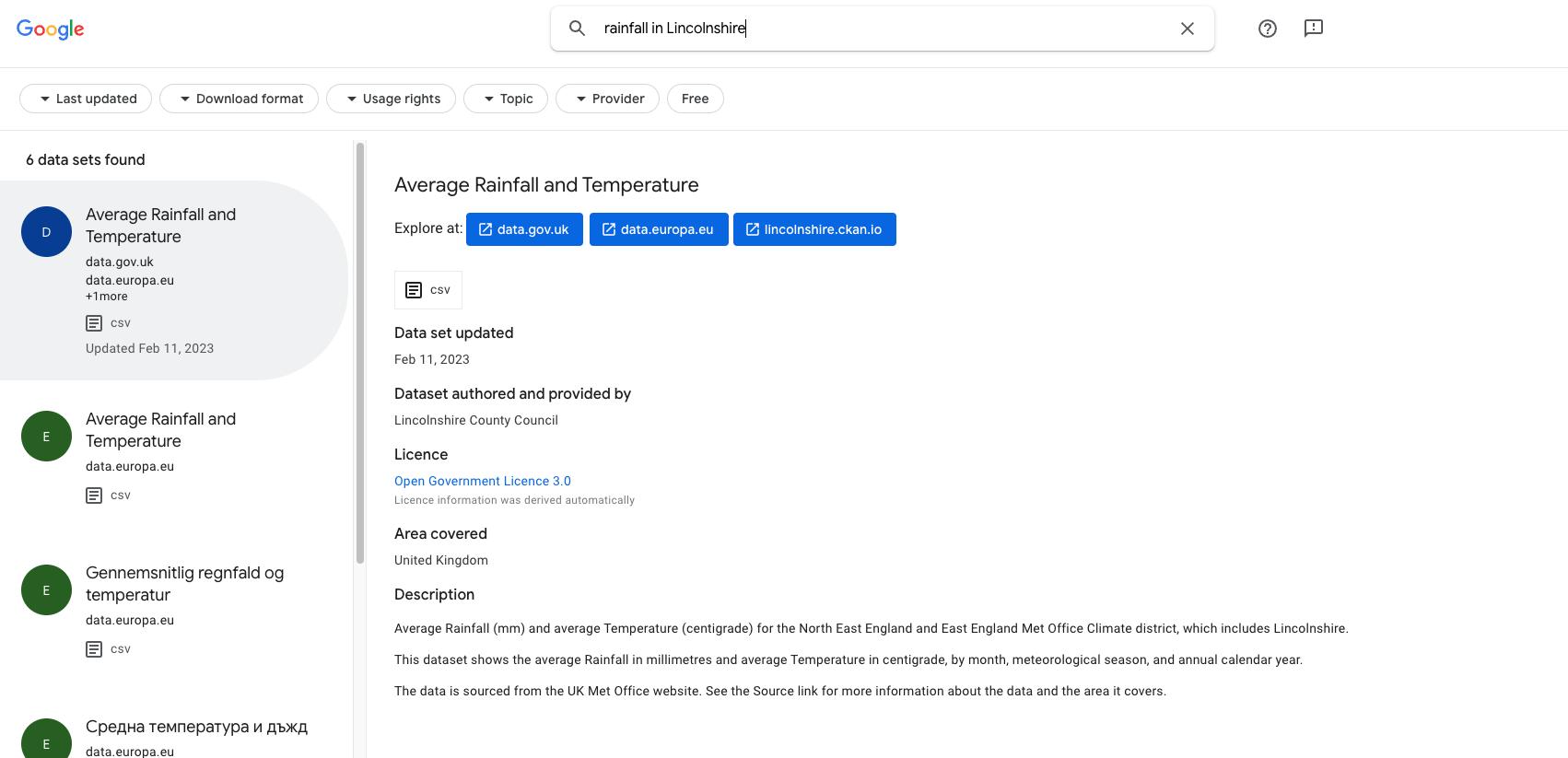}} 
    \subfigure[]{\includegraphics[width=0.4\textwidth]{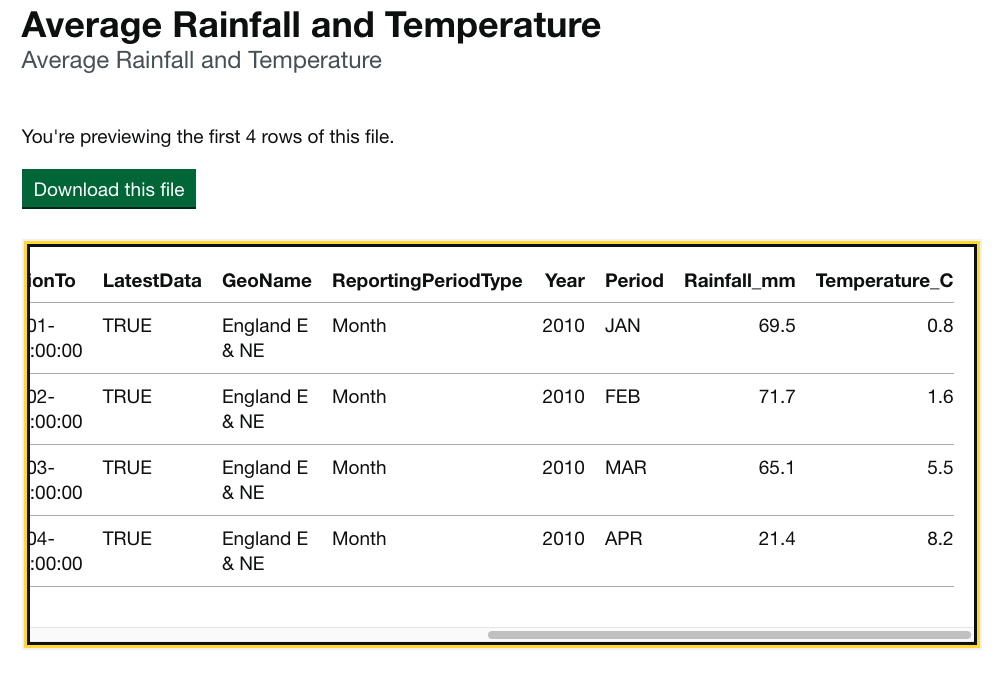}} 
    \subfigure[]{\includegraphics[width=0.45\textwidth]{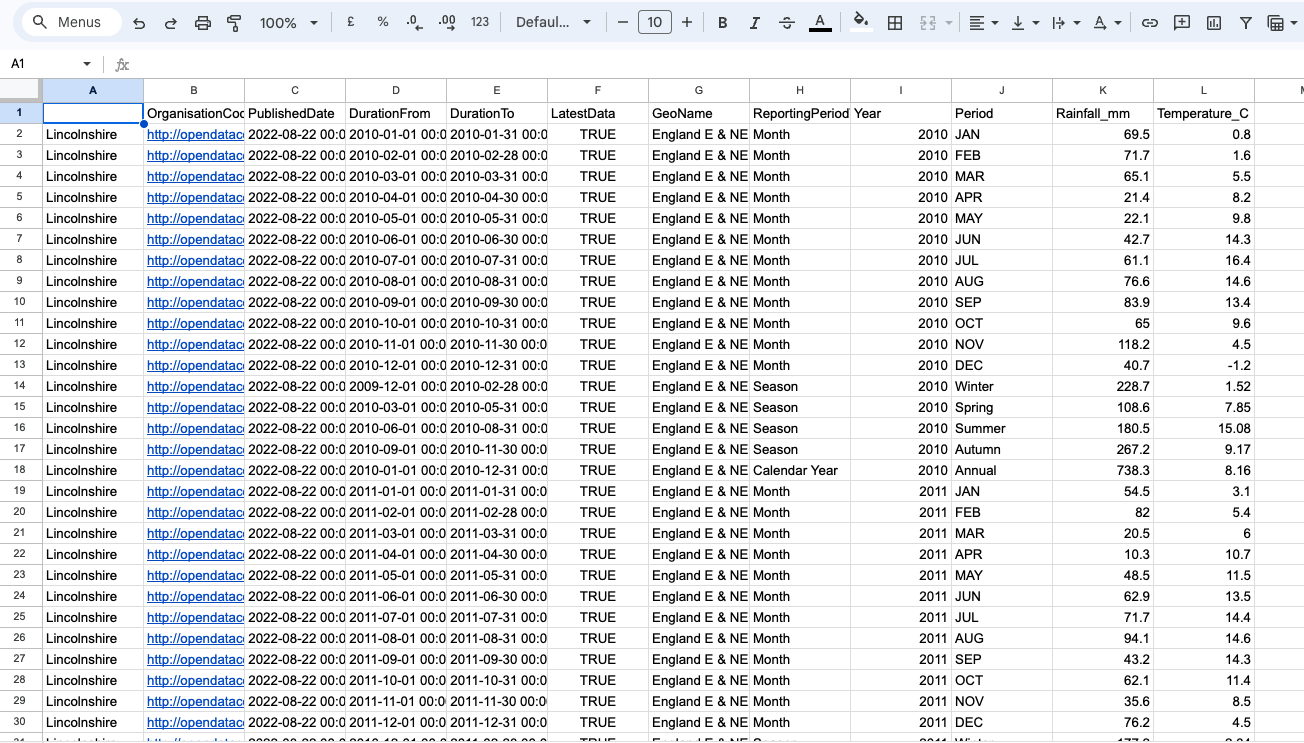}}
\subfigure[]{\includegraphics[width=0.4\textwidth]{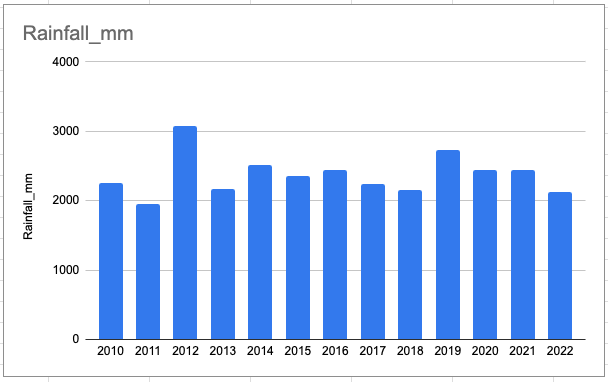}}
\subfigure[]{\includegraphics[width=0.45\textwidth]{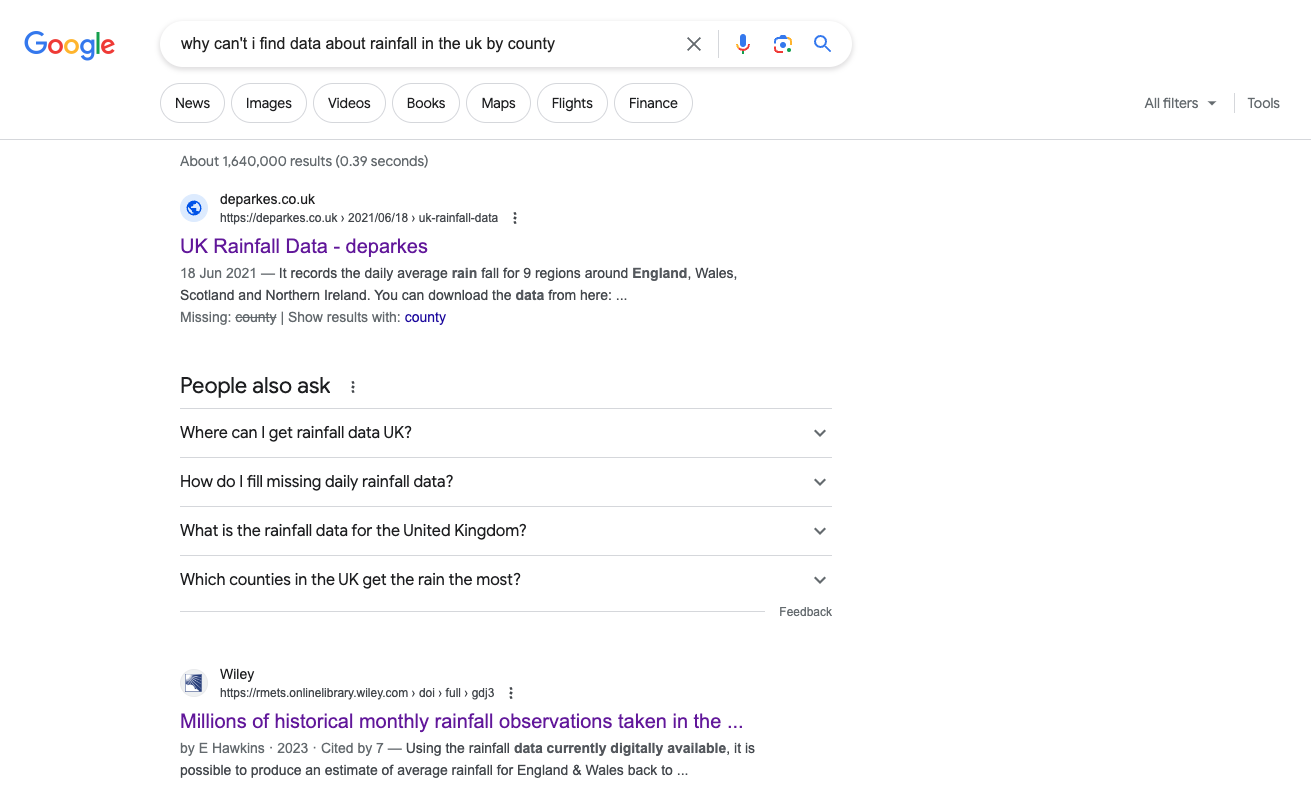}}
    \caption{Screenshots of a data discovery journey showing (a) dataset search on Google Data Search; (b) previewing the data on data.gov.uk website; (c) opening the data in excel; (d) data visualisation in excel and (e) evaluation information search on Google.}
    \label{fig:five_figures}
\end{figure}

While the user is able to find, explore and use the data, many issues remain. The user is not able to be explicit in their goal in undertaking the search, but only to best estimate the keywords that might locate data that aligns somewhat with the needs. Although the search tool offers summaries and previews, the user must engage with further tools to evaluate and explore the data, and then still more tools to actually employ it for the purpose for which it was sought. 

Further, data needs are complex, and can typically stretch over multiple search sessions, resulting in complex queries. These are poorly supported in web-based keyword search and can be better addressed with conversational search \cite{radlinski2017theoretical}.

Conversational generative artificial intelligence (CGAI) tools such as OpenAI's ChatGPT and Google's Bard have changed how we access information online \cite{hayes2023human}. We are therefore interested in exploring how they could change people’s interactions with data on the web. Given the ability of CGAI to summarise large, complex sources, explain results, and write code, we hypothesise that chatbots may also be able to find data matching a user goal and help the user understand its scope, quality, and provenance, and to use the data productively. This may also be supported further by the increased interactivity offered by conversational search, which has been established as empowering non-expert users \cite{mcmillan2002measures}. 

In this paper we explore whether, given a specific domain, are useful datasets suggested? Do the rationales make sense? Does CGAI help answer critical questions on data provenance, licensing, format and more, and how do users assess this? We begin to consider what a new generation of data tools, enabled by conversational generative AI, might offer to users, and how people might engage with them.  Our research questions, therefore, are:
\begin{itemize}
\item RQ1: How does CGAI support state of the art dataset discovery technologies?
\item RQ2: How do people express their information needs when using CGAI for data discovery?
\item RQ3. Which of the main sense-making data discovery activities previously identified are supplied by CGAI?
\end{itemize}
To address these questions we held workshops with users to explore how they would interact with CGAI when engaging in data discovery. The key contributions of this paper are a set of guidelines for data prompting with CGAI and a new prompting-centric human-data interaction framework.

\section{Background Literature}

In this section we contextualise our research in (1) the state of the art of dataset discovery and (2) CGAI for information retrieval.  

\subsection{Dataset discovery}

The manipulation, analysis, and sensemaking of data has been described as ‘data discovery’ \cite{chapman2020dataset}. The data we refer to here is structured (organised in tables, spreadsheets, databases and so on) and its potential scale is vast: as of 2023, Google Dataset Search indexed more than 45 million datasets from more than 13,000 websites \cite{GDS2023update}.  However, the data sought may not reside in a single dataset but across multiple sources, in a variety of formats, with no common ID, making sensemaking more complex \cite{elmqvist2011embodied}. 

\begin{figure*}[h]
\centering
\includegraphics[width=0.6\linewidth]{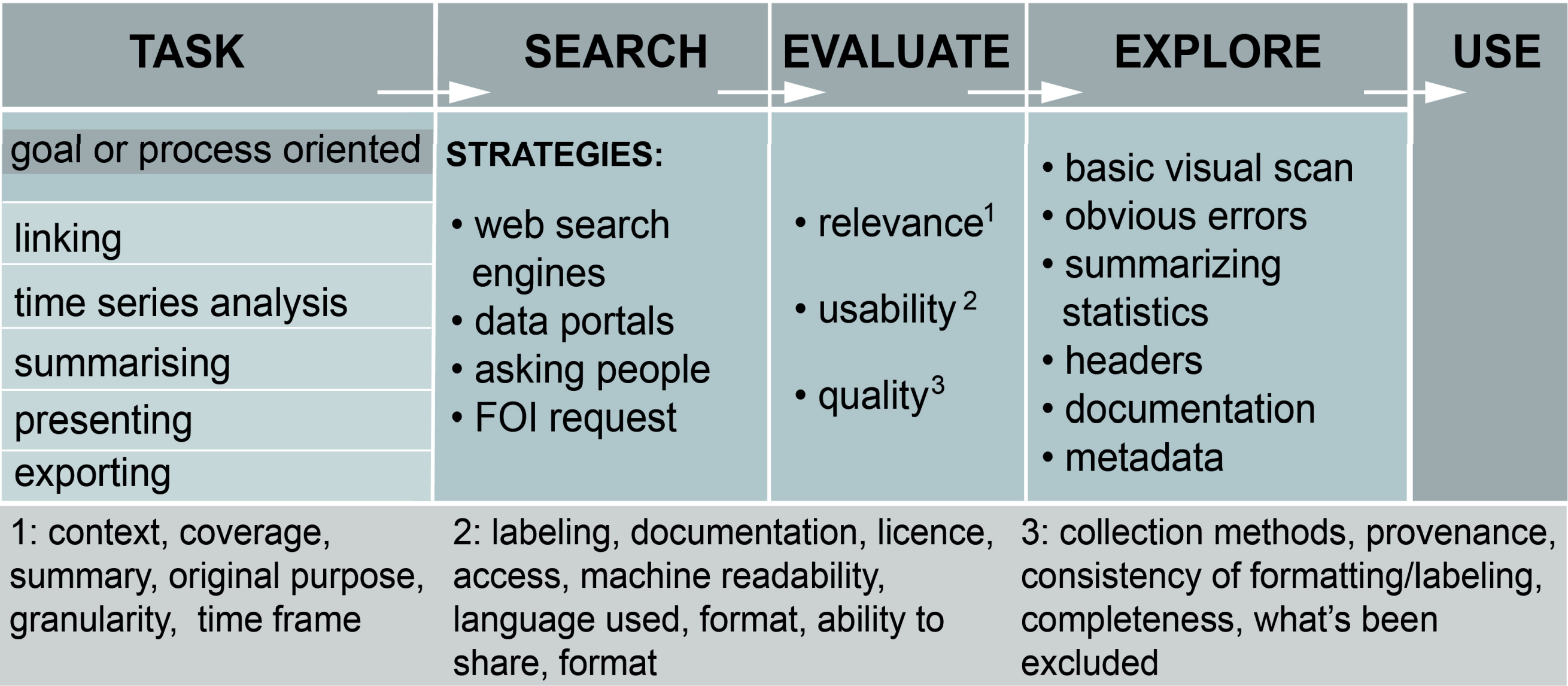}
\caption{Koesten et al's (2017) framework for interaction with structured data, showing the 5 steps of the data journey for users of secondary data. \cite{koesten2017trials}.}
\label{fig:Koesten_framework}
\end{figure*}

There are five key steps in the journey of interacting with structured data, which can be seen in Figure \ref{fig:Koesten_framework}, above \cite{koesten2017trials}. The first step of the data journey for the user is to define their motivation - why they want to use data. This task purpose is crucial to the process of selecting the appropriate dataset. Currently, there is no useful explicit way for users to define their task motivation in search, although search engines bring a wealth of implicit knowledge about previous user behaviour to the results they present. However, this excludes wider context about the task at hand \cite{shah2022situating}. 

The second step is that of search itself. Data search is a relatively recent field, with the first widely available dedicated dataset search on the web receiving a full launch as recently as January 2020 \cite{GDS2020launch}. The primary forms in which data is made available to people is via data repositories and portals (usually dedicated to a specific institution or domain); via the web, accessed by search engines; and through legal mechanisms such as Freedom of Information requests. When people need to look for data, they employ keyword or faceted search on portals or search engines. Additionally, people find data through personal recommendation \cite{ibanez2022comparison}. However, access to data does not equate enabling citizens to make use of the data \cite{vakulenko2020conversational}. ``There is a disconnect between what datasets are available, what dataset a user needs, and what datasets a user can actually find, trust and is able to use'' \cite{chapman2020dataset}. A solution to this challenge has not yet been satisfactorily implemented, although there have been improvements in supporting understanding of the data before downloading, such as providing visualisations and summarisation \cite{koesten2020everything}. 

Studies of online search behaviour pre-date the world wide web \cite{marchionini2006exploratory} and find that searches range from analytical, precise and carefully planned, to ``on the fly'' \cite{marchionini1988finding}. Users often find it challenging to describe their information needs because of a lack of background knowledge, and their understanding of an information need often evolves during the search process \cite{vakulenko2020conversational} \cite{radlinski2017theoretical}.  It has been suggested that when in need of information users actually want an domain expert, but turn to search engines in the absence of this \cite{metzler2021rethinking}. Unlike web searches for information, users searching for data are unlikely to expect an exact dataset match to their needs, but rather to locate a relevant subset of topical data \cite{kacprzak2017query}. Geospatial and temporal descriptions, constraints (eg open licence) and granularity are the most used parameters to specify data parameters in search \cite{kacprzak2019characterising}.

Data repositories aim to make data search easier, but come with their own complications: they tend to use metadata rather than keyword search \cite{koesten2017trials} and data users and publishers may use different terms for datasets. One study found that a search for ``housing stock'' on the UK government data website found almost twice as many datasets as a search for ``housing supply'' \cite{frank2016user}. Much of the time in learning search tasks is devoted to examining and comparing results and reformulating queries to discover the boundaries of meaning for key concepts \cite{marchionini2006exploratory}. Users also keep search terms short: the majority of searches on portals, both on the portal and via the web, are composed on average of less than 5 words \cite{kacprzak2019characterising}.

Further, the information retrieval task at hand may require an entire set of results from multiple different sources with interdependencies. A familiar example of this is booking a holiday, which requires a variety of interdependent travel and accommodation requirements to be met \cite{radlinski2017theoretical}. This kind of complex query requires ``memory'' of search interactions that is not yet met by the facilities of the web, and currently users searching for different elements of a holiday need to manually manage a range of concurrent searches which are not automatically informed by the parameters of each separate element. 

Sensemaking is a key condition for the secondary reuse of data and comprises the third and fourth steps of the data journey \cite{koesten2017trials,koesten2021talking}. Summary statistics or visualisations are often suggested as a way of exploring quantitative data, but this already implies a decision by the user that the data is worth engaging with, and that it has been assessed as fit for use \cite{brolchain2017extending,segel2010narrative,wang1998product}. To facilitate this, ‘immediate intelligibility’ is a key requirement for users when discovering data. However, without detail on how and why data was initially collected, this may present a challenge for even the most competent domain expert \cite{frank2016user}. These and other assessments of whether a dataset is fit for purpose, such as its granularity, provenance or format, have been described as ‘data evaluation’ \cite{koesten2017trials}. Such evaluations tend to be dataset and task specific rather than applicable to all datasets \cite{koesten2020everything,frank2016user}.  Users may need to make as many as 20 types of evaluations, related to the relevance, useability and quality of data \cite{koesten2017trials}. 

A second set of activities take place once the user begins to explore the data prior to use \cite{koesten2017trials}. These may include visualisations (provided or created by the user themselves) and a review of documentation, if available.  Text summaries are often used to reduce cognitive and time cost during information seeking activities online \cite{gambhir2017recent}. Research with dataset text summaries suggests that they do not necessarily contain all the required information (with methodology and provenance most frequently being ignored); that abstractive summaries require extensive natural language generation (NLG); and that they are subject to the idiosyncrasies of the people creating them, rather than being standardised \cite{koesten2020everything,gambhir2017recent}.Further, while it may be possible to generate a convincing summary of data itself, it may be the metadata or context that is more important for sensemaking \cite{wiseman2017challenges}. 

The multi-step complexity outlined above suggests that powerful NLG tools such as conversational search could be of great value to users requiring data. Conversational search can build understanding and semantic coherence over multiple rounds of interaction, while also supporting user needs through clarification and suggestion, \cite{radlinski2017theoretical,zhang2018towards}. Conversational search is particularly useful for exploratory information-seeking contexts where users have insufficient background knowledge \cite{schneider2023investigating}. While there is a substantial corpus of research on this, pre-dating the existence of large language models, there is little to no research on conversational dataset search \cite{schneider2023investigating,vakulenko2020conversational}.

\subsection{CGAI for information retrieval}

Generative AI refers to a class of artificial intelligence models that can create new data based on patterns and structures learned from existing data. Recently there has been explosive growth in conversational agents based on Large Language Models (LLMs). Such conversational generative artificial intelligence (CGAI) can be used to address natural language tasks,  commonly include writing essays or literature reviews of scientific papers, and reading and summarising reports \cite{salvagno2023can, alkaissi2023artificial, sallam2023chatgpt}. They can also be used to answer complex natural language questions and support the answers with justifications. Further, many LLMs operate in multiple spoken languages and can therefore both translate text and justify the constructions used in translation, as well as providing text in many coding languages \cite{mialon2023augmented}. CGAI has been used in a number of text manipulation and generation tasks such as creating financial research , structuring unstructured information and detecting stance in social media posts \cite{dowling2023chatgpt, wei2023zero, aiyappa2023can}. It has been found to outperform crowd workers on text annotation tasks such as identifying types of social media posts \cite{gilardi2023chatgpt}.

In conversational search powered by LLMs, users are not restricted to keywords, and are able to state requests in a number of different ways. This suggests they can include more description about their information need in their questions. However, while this may increase the information available to the model to generate an appropriate answer, misleading or unnecessary information may adversely affect results \cite{zuccon2023dr}. The solution to this is generally held to lie in prompt engineering; designing the requests in such a way that they minimise error and maximise potential for appropriate results \cite{bowman2023eight}. This requires a certain level of awareness of potential errors, skill in devising prompts to reduce such errors and assessing the effectiveness of prompts, which is challenging to non-experts \cite{zamfirescu2023johnny}.  Prompts can also be used to drive the LLM to initiate question asking, by presenting a task that requires the agent to discover more information from the user to complete it \cite{white2023prompt}. Research into using GPT to extract unstructured to structured information has shown that greater success is achieved when the information extraction task is broken down into easier sub-tasks, which again requires specific skills \cite{wei2023zero}. A further strategy for improving search is to specify a domain-expert persona for the LLM to adopt. However, this has been found to also increase bias in language tasks \cite{salewski2023context}. 

LLMs do not have access to contemporaneous, proprietary or personal data. Consequently, enabling access to external data sources, and augmenting CGAI for retrieval of information, in particular through detecting and calling APIs, is a major area of research \cite{halevy2023learnings, jin2023genegpt}. Another is addressing the issue of hallucinations.  A hallucination is where a plausible answer returned is not based on existing fact, and is ‘made up’ \cite{aiyappa2023can}. Hallucinations are of particular interest in our research as they often occur in the form of references to sources that do not actually exist \cite{alkaissi2023artificial}. The ability to identify hallucinations presupposes a certain amount of pre-existing knowledge on the part of the user, as LLMs are unable to  directly refer to supporting documents in their training data \cite{metzler2021rethinking}  There is currently no reliable way for developers to  steer the behaviour of LLMs, and even less so by inexperienced users \cite{bowman2023eight, zamfirescu2023johnny}. The ability of LLMs to fluently produce apparently coherent text may mask the fact that this text may be of much lesser quality than it first appears (containing very little truth or fact), or that it may accidentally be damaging \cite{bender2021dangers}. 

Explaination is desirable (and in some cases, legally required) in intelligent systems. It makes it possible for users to act appropriately, is important for discovering and understanding novel concepts, and supports users' trust in an application \cite{hoffman2023increasing, confalonieri2021historical}. A working definition of explainable AI (XAI) is ''Given an audience, an explainable Artificial Intelligence is one that produces details or reasons to make its functioning clear or easy to understand'' \cite{arrieta2020explainable}. Explanations have some basic requirements, but they are largely varied based on the requirements, background knowledge, and so forth of the audience. Explanations are used to justify decisions, control function, improve accuracy and discover new knowledge \cite{vilone2020explainable}. There are as many as $36$ various notions of explanations, including efficiency (whether the explanation supports faster decision making), faithfulness (whether definitions truly reflect the datapoints) and robustness (the consistency of the results of the method \cite{vilone2020explainable}.

 Explanations have a social dimension - they constitute an element of a dialogue that aims to transfer knowledge, therefore engaging the beliefs of the giver and recipient of the explanation \cite{vilone2020explainable}. There is concern that explainability is dealt with primarily technically and less attention is spent on whether the explainability system is effective and efficiently human-centred \cite{mueller2021principles}. While much attention is currently being paid to understanding how models such as ChatGPT work, much of this insight is currently speculative \cite{zamfirescu2023johnny}. Consequently, most explanations for LLMs are post-hoc (the dataset, training methods and algorithms are opaque) and most focus on single data points, rather than the whole model.  It is, of course, theoretically possible for a user to request that ChatGPT or other model ‘reflects’ on its answer and explains how it arrived at it \cite{white2023prompt}.

\subsection{Summary }

People both need to locate and understand datasets in order to reuse them. This is often done as part of work, both paid and volunteer. This requires skills around crafting appropriate search, looking in appropriate locations and then assessing the fitness for use of the data. Despite improvement on the supply side, there is still a huge time and cognitive cost on the demand side. Conversational generative artificial intelligence (CGAI) tools such as ChatGPT and Bard have changed how we access information online \cite{sallam2023chatgpt}. We are therefore interested in exploring how they could change people’s interactions with data on the web. Given the ability of CGAI to summarise large, complex sources, explain results, and write code, we hypothesise that chatbots may also be able to find data matching a user goal and help the user understand its scope, quality, and provenance, and to use the data productively. In the next section we describe our user studies with conversational dataset search.
.

\section{Methodology}

To explore how users might engage in conversational dataset search, we invited a gender-balanced group of users to three workshops conducted in two languages, in which we asked them, ``to think of one dataset (or group of datasets) that you would like to discover, using [CGAI agent]. (This can be a dataset you know exists, or a completely new-to-you dataset.) Once [agent] has found something (this may require more than one prompt), explore what it has returned to see if it is usable''. For further details see Appendix \ref{appendix:a}. 

\subsection{Data collection}

 Our pilot was run using the dominant large language model-based chatbot at the time, OpenAI's GPT3.5. We wished to replicate the scenario we felt was most likely, which was to use the CGAI which had most public awareness and users. For our second workshop, we tried to account for some of the limitations of GPT3.5 by using GPT4. However, 2 days before our research, the live web search facility of GPT4 was suspended \cite{gpt2023disconnect}.  In our third workshop, therefore, we used an (easily accessible) LLM chatbot with web connectivity. As Microsoft Bing is built upon similar foundations to ChatGPT, we opted for Google Bard. Details of the workshops are shown in Table \ref{tab:workshop_details}.

\begin{table}[h]
\caption{Overview of workshop details.}
\label{tab:workshop_details}
\begin{tabular}{|p{0.01\linewidth}|p{0.08\linewidth}|p{0.14\linewidth}|p{0.15\linewidth}|p{0.01\linewidth}|p{0.09\linewidth}|p{0.14\linewidth}|p{0.08\linewidth}|p{0.08\linewidth}|}
\hline
\textbf{\#} & \textbf{Date} & \textbf{CGAI}& \textbf{Rationale} & \textbf{n} & \textbf{Gender} & \textbf{Attendees}  & \textbf{Location} & \textbf{Language} \\ \hline
1           & 22/05/2023    & Chat GPT3.5 & Highest adoption   & 9          & m=4, f=5        & AI and LLM experts  & London            & English           \\ \hline
2           & 07/07/2023    & Chat GPT4 & Improved performance on GPT3.5    & 5          & m=3, f=1, e=1   & Political activists & Berlin            & German            \\ \hline
3           & 12/07/2023    & BARD  & Web connectivity        & 9          & m=4, f=5        & Data professionals  & London            & English           \\ \hline
\end{tabular}
\end{table}

The pilot workshop was conducted in London, UK with users with PhD level understanding of AI and large language models. This was to confirm the format of the workshop and protocol. Eight participants took part in the workshop in person, and 2 trialled a self-guided, asynchronous process following the protocol closely. After this workshop we made two adjustments. Firstly we decided to audio record the ``in the room'' discussion as this provided rich data and background that supported the data in the questionnaires. Secondly, we found the asynchronous process yielded substantially less research data so abandoned this mode. We used the ChatGPT web interface at chat.openai.com, which does not allow any special features or the uploading of large amounts of text or changes of hyperparameters such as temperature settings \cite{zuccon2023dr}. 

For our following workshops, we aimed to ask non-specialist users in a real world context to define a problem they have, and then think about what data they could use to address it. As it was important to our research to reflect a real world scenario, we aimed to work with people who were staff or activists in resource-constrained organisations, who would strongly benefit from using data and understanding its accuracy and provenance. We accessed these groups via personal networks.

Our second workshop was held in Berlin, Germany with professionals who were engaged in political activism. This used the ChatGPT4 web interface, in order to benefit from any relevant improved performance, for instance the ability to input longer texts. All participants were present in person. The third workshop, in London, comprised data professionals with familiarity with the data economy and how data is used, but whose roles did not involve active coding. Three participants were located remotely but engaged in the workshop process synchronously via Teams. The number of chats and questionnaires collected are shown in Table \ref{tab:data}.

\begin{table}[h]
\caption{Research data collection.}
\label{tab:data}
\begin{tabular}{|l|l|l|l|l|l|l|}
\hline
\textbf{Date}  & \textbf{Location} & \textbf{CGAI} & \textbf{Questionnaires} & \textbf{Conversations} & \textbf{Audio} & \textbf{Teams} \\ \hline
18/05            & London             & ChatGPT3.5    & 9                       & 9                           & N/A            & N/A            \\ \hline
7/07           & Berlin            & ChatGPT4      & 5                       & 10                          & Yes            &                \\ \hline
12/07          & London            & Bard          & 9                       & 8                           & Yes            & Yes            \\ \hline
\textbf{Total} & 4                 & 3             & 23                      & 27                          & 2              & 1              \\ \hline
\end{tabular}
\end{table}

During the workshop, participants were provided with the guidance that can be found in Table \ref{tab:guidance}, which resulted in the associated output (a full version of the guidance can be found in Appendix \ref{appendix:a}):

\begin{table}[h]
\caption{Activity guidance and associated output from the workshops.}
\scriptsize
\label{tab:guidance}
\begin{tabular}{|p{0.7\linewidth}|p{0.25\linewidth}|}
\hline
\textbf{Guidance}  & \textbf{Output}                                \\ \hline
\begin{tabular}[c]{@{}l@{}}Think about your motivation for a dataset search. \\ Enter your motivation and any expectations (e.g. I know this dataset really well and know what I expect to see/I don’t know\\ if anything like this exists) into the questionnaire\\ Devise your prompt for {[}agent{]}\\ Enter any considerations you had when devising your prompt into the questionnaire\end{tabular} & \begin{tabular}[c]{@{}l@{}}Audio discussion (for workshops 2 and 3). \\ Questionnaire q1-7 {[}Appendix \ref{appendix:b}\end{tabular}  \\ \hline
Enter your prompt and explore {[}the agent’s{]} response                      & Conversational transcript  \\ \hline
As a group, discuss ChatGPTs response and your assessment of this. Enter replies to the questionnaire as we go.                                                                                                          & \begin{tabular}[c]{@{}l@{}}Audio discussion (for workshops 2 and 3). \\ Questionnaire q8-11 {[}Appendix \ref{appendix:b}{]}\end{tabular} \\ \hline
\end{tabular}
\end{table}

We asked participants to evaluate their own experience of using a conversational agent as Successful, Somewhat successful or Not successful. We then broke this down into two elements. We asked whether an appropriate dataset had been suggested (locating data) and then we asked participants to indicate on a Likert scale how useful the supporting information provided with the dataset was (sense-making). Finally we asked participants to compare the process with previous search experience, if relevant. 

\subsection{Data preparation and analysis}

The datasets, analytical methods applied and the relevant research questions are shown in Table \ref{tab:RQ_data}.

\begin{table}[h]
\caption{Data collection and analysis for each of the research questions addressed. The code for the descriptive statistics is available as a zip file.}
\label{tab:RQ_data}
\begin{tabular}{|p{0.3\linewidth}|p{0.35\linewidth}|p{0.25\linewidth}|}
\hline
\textbf{RQ}                                                                                                          & \textbf{Data}                                                                                                                    & \textbf{Analysis}                                                                             \\ \hline
1. How can conversational agents support SOTA dataset discovery technologies?                                        & \begin{tabular}[c]{@{}l@{}}Questionnaire questions 1, 2, 7, 8, 9\\ Chat sessions\end{tabular}                                    & \begin{tabular}[c]{@{}l@{}}Descriptive statistics \\ Inductive thematic analysis\end{tabular} \\ \hline
2: How do people express their information needs?                                                                    & \begin{tabular}[c]{@{}l@{}}Questionnaire demographics\\ Chat sessions\\ Questionnaire questions 3, 4, 5, 10, 11, 12\end{tabular} & \begin{tabular}[c]{@{}l@{}}Descriptive statistics\\ Inductive thematic analysis\end{tabular}  \\ \hline
3. Which of the main sense-making data discovery activities previously identified are supplied by Conversational AI? & \begin{tabular}[c]{@{}l@{}}Chat sessions\\ Questionnaire questions 3, 4, 5, 10, 11, 12\end{tabular}                              & Deductive thematic analysis based on Koesten et al (2017)                                     \\ \hline
\end{tabular}
\end{table}

The data was prepared as follows.
The audio recordings were uploaded onto our institutional OneDrive, transcribed and (where necessary) translated using the OneDrive translate facility. The chat sessions and questionnaires were uploaded into our institutional OneDrive and again translated using the OneDrive translate facility. All translations were checked by a German speaker for accuracy. 

The audio transcriptions, questionnaires, and chat session files were then uploaded into NVIVO for thematic coding by two of the researchers. Two of the authors individually coded a sample of the chat transcripts and long form questionnaire answers inductively for RQ1 and RQ2, and coded a sample of the chat transcripts deductively for RQ3. The resulting code lists were compared and conflicting results were discussed. This was repeated twice for both the deductive codes and inductive codes.  The resulting code books are attached as Appendix \ref{appendix:c}. The short questions were analysed quantitatively using Python, and the code is available as a zip file. We counted the frequency of responses and their combinations. We calculated the length of the chat sessions by counting the number of queries asked, the responses when interacting with the conversational agent, and their number of words.

\subsubsection{Ethics}
The interview study was approved by Author 1's  institution’s Ethical Advisory Committee via the Minimal Risk Procedure. Informed verbal (audio recordings and chat transcripts) and written (questionnaire) consent was given by the participants. No personal information was used in the analysis. 

\section {Findings}

\subsection{RQ1: How do CGAI compare with state of the art dataset discovery technology?}

In this section we examine whether conversational search can compete in terms of success and usefulness for dataset discovery. We explore how people are using it in terms of the length of their queries and length of chats, and also examine some aspects of explanations. 

\begin{figure*}[t]
\centering
\includegraphics[width=0.6\linewidth]{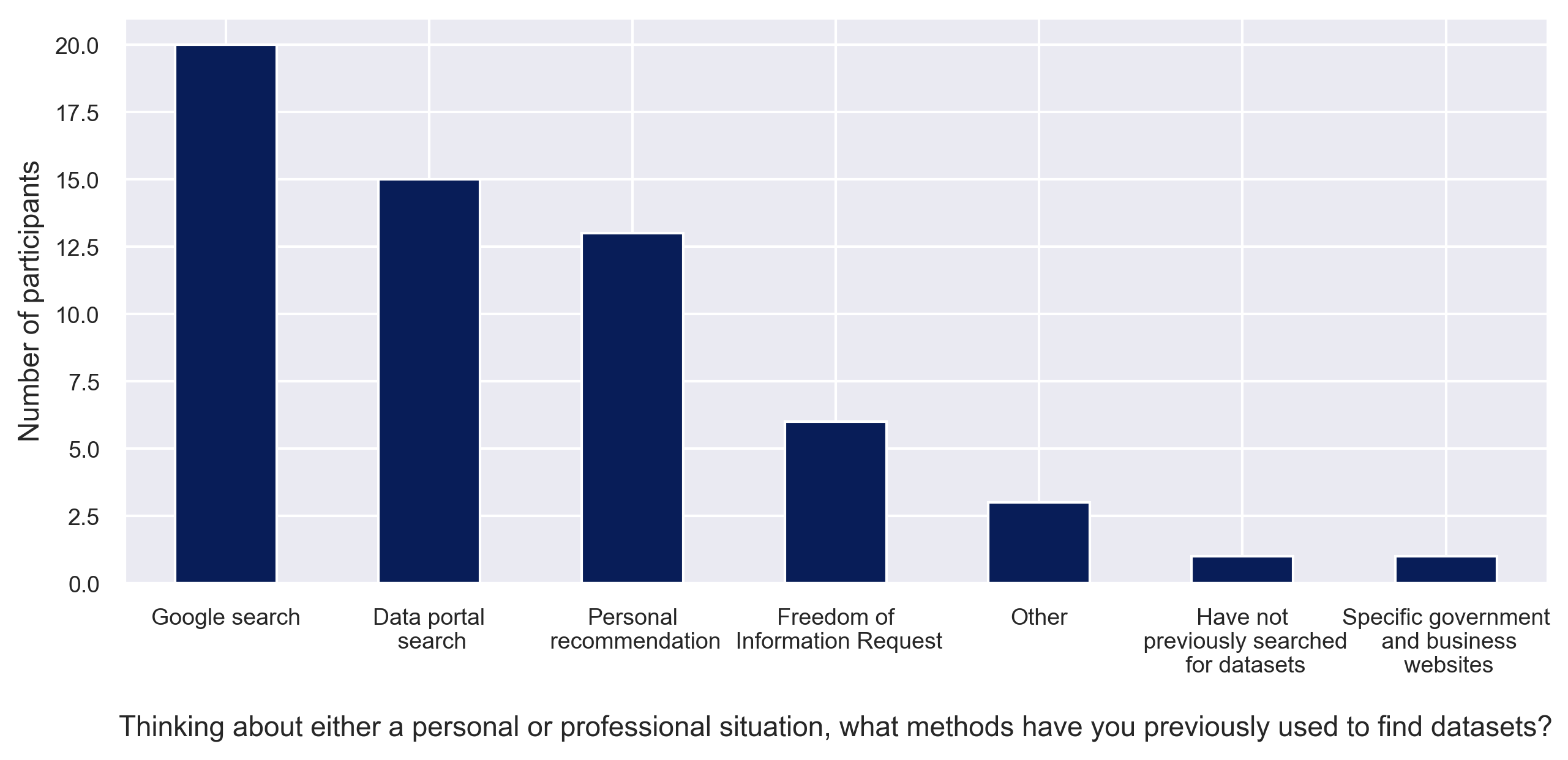}
\caption{Results from the questionnaires regarding previously used methods (multiple answers possible).}
\label{fig:other_search}
\end{figure*}

\begin{figure*}[h]
\centering
\includegraphics[width=0.5\linewidth]{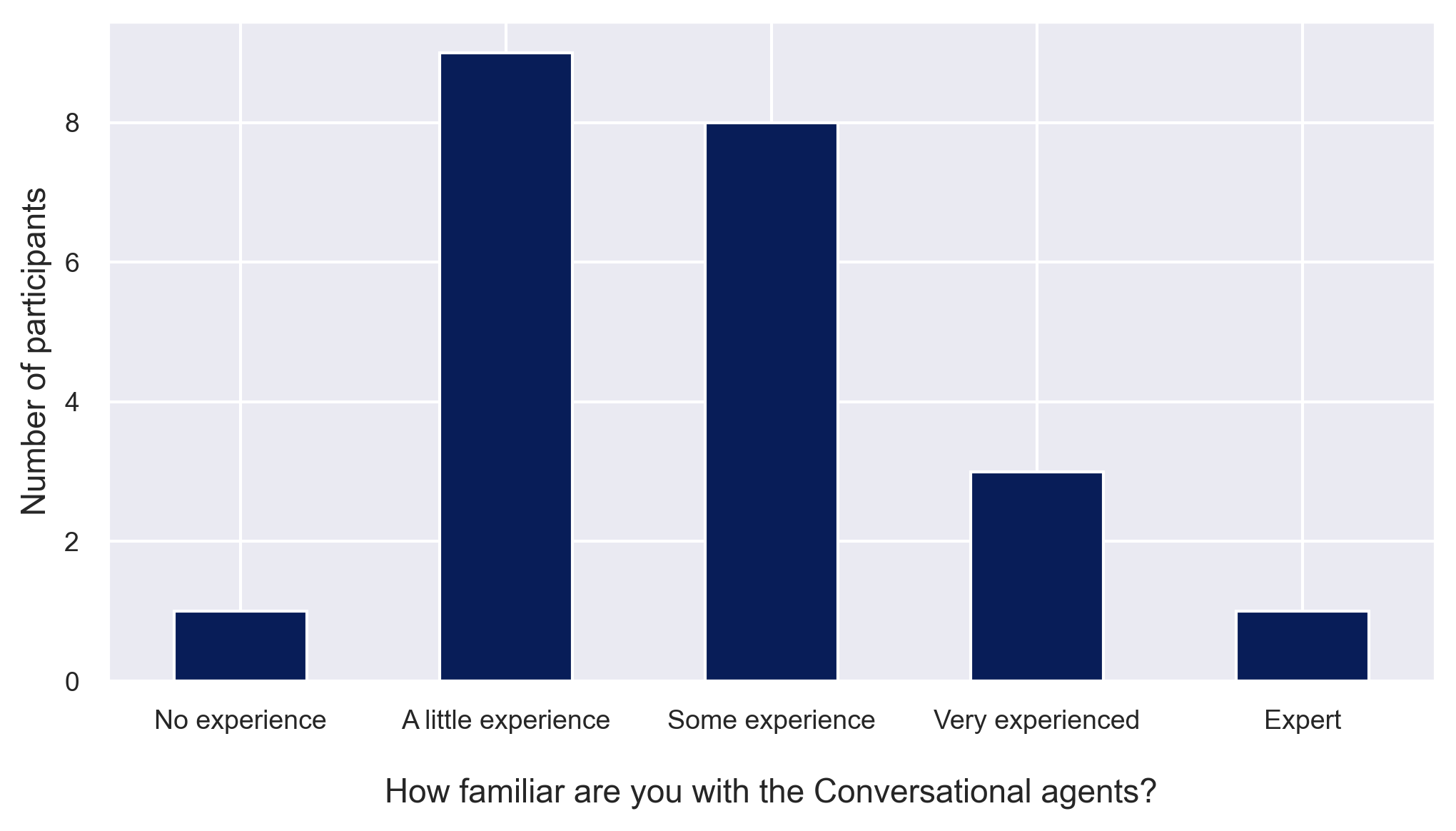}
\caption{Results from the questionnaires regarding familiarity with the conversational agents.}
\label{fig:familiar}
\end{figure*}

\subsubsection{Success rates of conversational dataset search}
\begin{figure}[h]
    \centering
    \subfigure[]{\includegraphics[width=0.45\textwidth]{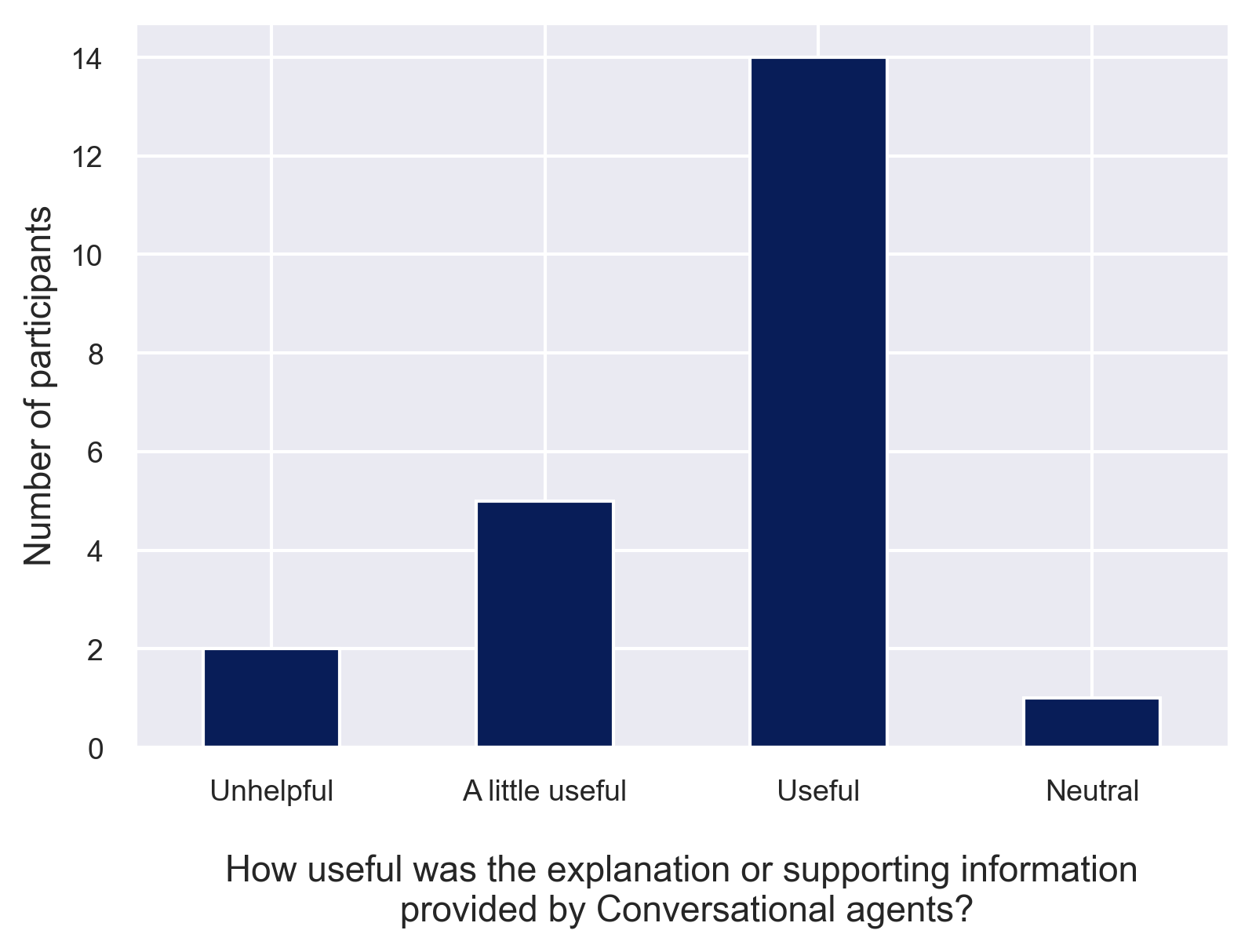}} 
    \subfigure[]{\includegraphics[width=0.4\textwidth]{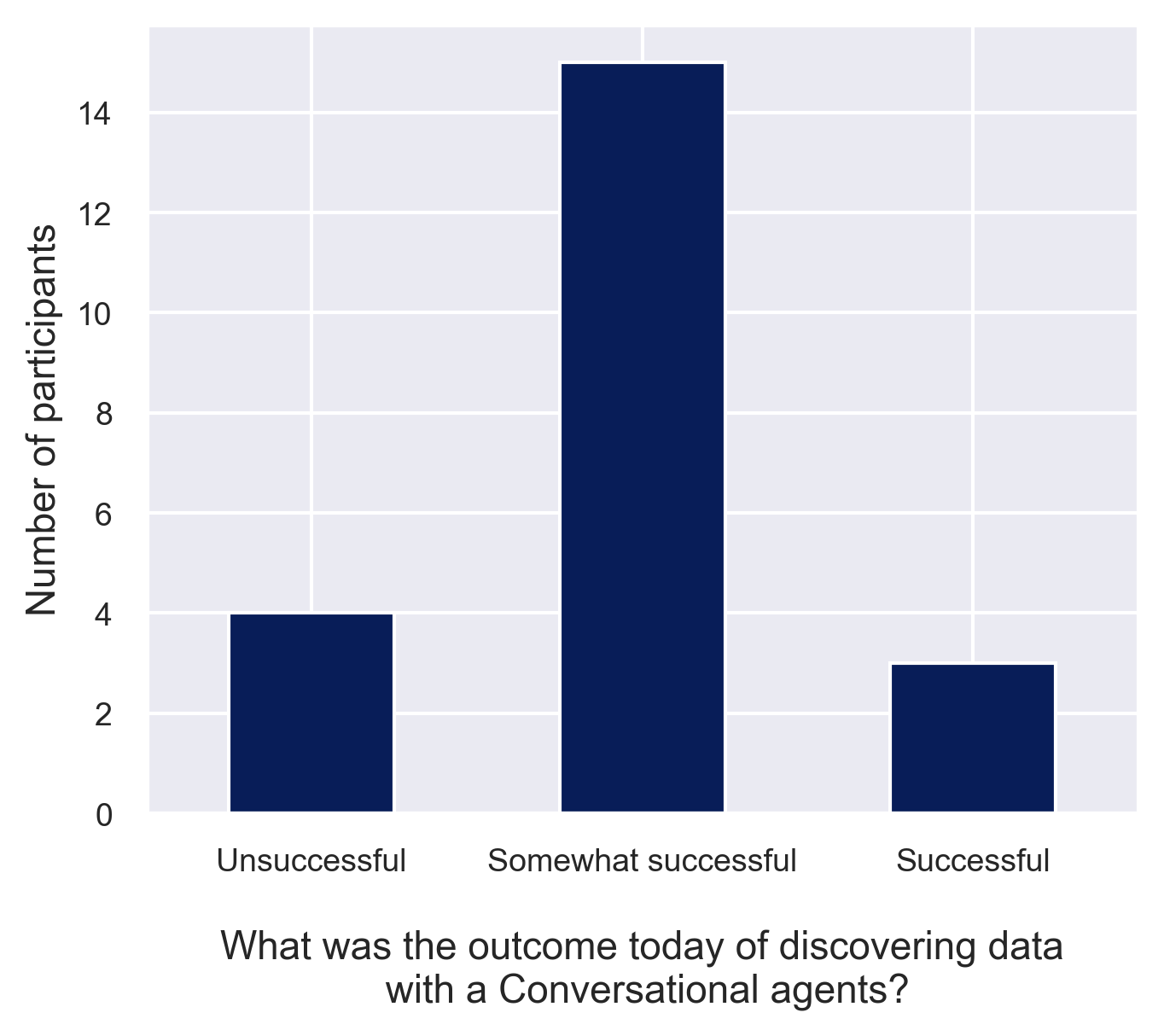}}
    \caption{The figures present results from the questionnaires after the chat with the conversation agents regarding (a) how useful was the support of the agent (b) how successful was the outcome.}
    \label{fig:useful_success}
\end{figure}

In order to assess our participants' previous experience with datset search, we asked them what methods they have previously used to find datasets. While one participant had not previously searched for datasets at all, most of the participants had used Google search and specific data portals (Figure \ref{fig:other_search}). In addition, we asked participants how familiar they were with the conversational agents (Figure \ref{fig:familiar} ). The majority had some or a little experience with them. Only one particpant had no experience.
After the conversation with the chatbot, we asked participants about the process (Figure \ref{fig:useful_success}). The majority of participants (n=15) found the process of using chat agents ``somewhat successful''. It is interesting to note that despite a large number of participants finding the process overall ``useful'' (n=14), only three of them found it fully ``successful'' in terms of finding a dataset. 


\subsubsection{Query and chat format}

\begin{figure}[h]
  \centering
  \includegraphics[width=0.8\textwidth]{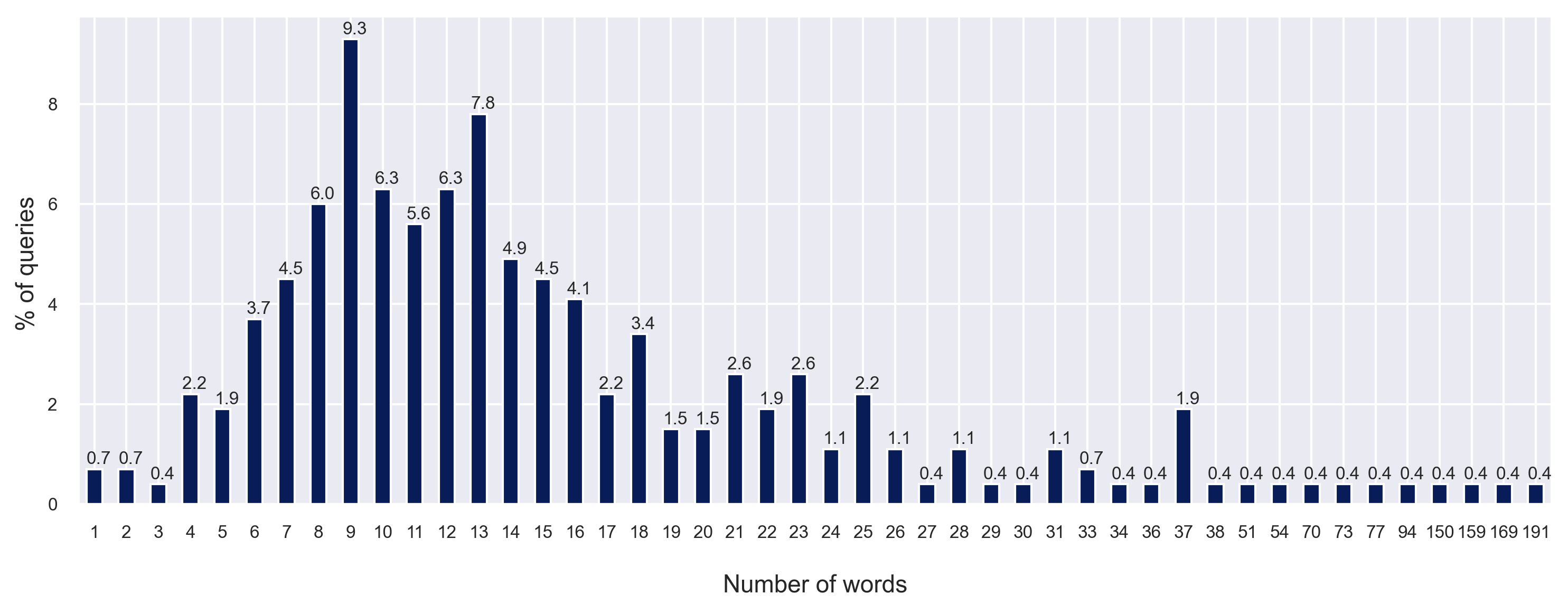}
  \caption{ The figure presents an overview of the length of queries in terms of the number of words.}
  \label{fig:num_word_queries}
\end{figure}

As can be seen in Figure \ref{fig:num_word_queries}, the total number of words in a query had a wide range, from $1$ - $191$. The modal number was $9$.

\begin{figure*}[h]
\centering
\includegraphics[width=0.5\linewidth]{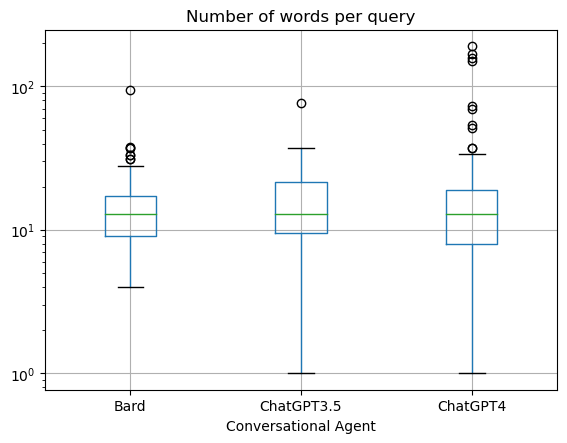}
\caption{The box plot shows the range of query length for the three conversational agents.}
\label{fig:agents_box}
\end{figure*}

We found that the mean number of words per query is approximately 18. This differs from data portals and web search, where the average number of words per query on a data portal is fewer than $3$, while using the web search and then redirecting to a data portal is fewer than $5$ \cite{kacprzak2017query}. However, it reflects the increased average length of prompts for text-to-image generation for which the mean and median have been found to be above $20$ words \cite{xie2023prompt}.Figure \ref{fig:agents_box} shows that while the median number of words per query is similar across all three agents, participants using Bard did not use as many long queries.

In terms of the queries themselves, we collected $268$ queries across $27$ chats with conversational agents. Previous research on conversational search using a text chat interface received $669$ queries for $26$ chats \cite{schneider2023investigating}. However, this study used human participants playing the part of the conversational agents, which may have affected the length of chat. In addition, we found great variation in the range of queries per chat. Our study includes chats with $1$ to $33$ queries and on average $10$ queries per chat as shown in Table \ref{tab:stats_agents}. The study above also found minimum and maximum message count varied widely, from  $8$ messages to $62$ messages, and an average $26$ queries per chat. We found our average number of queries was consistent across our German speaking group and our English users as shown in  Table \ref{tab:stats_language}.

\begin{table}[h]
\caption{The table presents descriptive statistics for the queries with the conversational agents.}
\label{tab:stats_agents}
\begin{tabular}{|l|l|l|l|l|}
\hline
\textbf{Statistics}                & \textbf{ChatGPT4} & \textbf{ChatGPT3.5} & \textbf{Bard} & \textbf{Total} \\ \hline
Number of chats                    & 10                & 9                   & 8             & 27             \\ \hline
Number of queries                  & 105               & 55                  & 108           & 268            \\ \hline
Min-Max queries per chat           & 1-23              & 1-14                & 3-33          & 1-33           \\ \hline
Average number of queries per chat & 10.5              & 6.1                 & 13.5          & 9.9            \\ \hline
Average number of words per query  & 22.9              & 18                  & 17            & 17.7           \\ \hline
Average number of words per answer & 185.2             & 214.1               & 349.8         & 215.3          \\ \hline
\end{tabular}
\end{table}

\begin{table}[h]
\caption{The table presents descriptive statistics for the queries with the conversational agents in different languages.}
\label{tab:stats_language}
\begin{tabular}{|l|l|l|}
\hline
\textbf{Statistics}                & \textbf{English} & \textbf{German} \\ \hline
Number of chats                    & 17               & 10              \\ \hline
Number of queries                  & 163              & 105             \\ \hline
Min-Max queries per chat           & 1-33             & 1-23            \\ \hline
Average number of queries per chat & 9.6              & 10.5            \\ \hline
Average number of words per query  & 17.6             & 22.9            \\ \hline
Average number of words per answer & 278              & 185.2           \\ \hline
\end{tabular}
\end{table}

\subsubsection {Results of data requests}

\begin{table}[]
\caption{Table showing number of datasets (or close sources, such as portals) returned after first request. A ``request'' dataset is named, a ``described'' dataset is defined but not named and an ``implied'' dataset asks for data as information, but does not specify it explicitly as data. Comment is appended by the authors. }
\scriptsize
\label{tab:dataset_search}
\begin{tabular}{|p{0.12\linewidth}|p{0.09\linewidth}|p{0.05\linewidth}|p{0.2\linewidth}|p{0.05\linewidth}|p{0.3\linewidth}|}
\hline
\textbf{Data}     & \textbf{Request}         & \textbf{LLM} & \textbf{Response}                                                   & \textbf{Datasets} & \textbf{Comment}                                                                \\ \hline
Case law          & Implied                  & GPT4         & No access to database                                               & 1                 & curated micro dataset                                                           \\ \hline
Case law          & Implied                  & GPT4         & Does not mention having no access to database                       & 1                 & curated micro dataset                                                           \\ \hline
Open source       & Described                & GPT4         & No access to specific datasets                                      & 0                 & Suggested 4 strategies for search. (on being prompted for sources, suggested 5) \\ \hline
Birds             & Described                & GPT4         & No excuses                                                          & 4                 &                                                                                 \\ \hline
Birds             & Described                & GPT4         & Does not have access to datasets, also problem with internal server & 0                 & 5 questions later it suggests 3, although not the same as those above           \\ \hline
Plug ins          & Described                & GPT4         & Not up to date                                                      & 0                 & Recommends contacting Open AI                                                    \\ \hline
Addresses         & Described                & GPT4         & Not up to date, no access to db                                     & 5                 &                                                                                 \\ \hline
Complex numbers   & Mathematical explanation & GPT4         &                                                                     &                   &                                                                                 \\ \hline
Complex numbers   & Mathematical explanation & GPT4         &                                                                     &                   &                                                                                 \\ \hline
Student numbers   & Implied                  & GPT4         & No access to database                                               & 0                 & Recommended where to contact                                                    \\ \hline
Property prices   & Implied                  & GPT3.5       & No real time access                                                 & -                 & Text summary describing the data                                                \\ \hline
Ontology          & Described                & GPT3.5       &                                                                     & 7                 & With supporting information on what an ontology is                              \\ \hline
Conferences       & Described                & GPT3.5       & Yes!                                                                & 15                &                                                                                 \\ \hline
Knowledge graphs  & Described                & GPT3.5       & Yes                                                                 & 5                 &                                                                                 \\ \hline
Credit scores     & Requested                & GPT3.5       & Yes                                                                 & 1                 & With link                                                                       \\ \hline
Islamophobia      & Described                & GPT3.5       & No                                                                  & 0              & Provides some later                                                             \\ \hline
Music emotion     & Described                & GPT3.5       & Sure!                                                               & 8                 &                                                                                 \\ \hline
Panama paper      & Request                  & GPT3.5       & Not able to access                                                  & 1                 & With supporting information                                                     \\ \hline
Fire hydrants     & Described                & Bard         & Sure                                                                & 1                 & partial dataset, states it is fictional                                         \\ \hline
Trees             & Described                & Bard         & Yes                                                                 & 1                 & With link and usage                                                             \\ \hline
Longitudinal data & Described                & Bard         &                                                                     & 6                 & ‘Most popular’ with links                                                       \\ \hline
Sport             & Described                & Bard         &                                                                     & 5                 & With links plus suggestions for more                                            \\ \hline
Homeslessness     & Described                & Bard         & Sure!                                                               & 1                 & Curated dataset from unascribed source and some interpretation of trend         \\ \hline
Physics           & Implied                  & Bard         &                                                                     & 0                 & Describes interpretation of underlying data                                     \\ \hline
Junk food         & Implied                  & Bard         &                                                                     & -                 & Textual summary of report                                                       \\ \hline
Peer production   & Described                & GPT3.5       &                                                                     & 3                 & In textual report form rather than list                                         \\ \hline
Education         & Implied                  & Bard         &                                                                     & -                 & Textual report                                                                  \\ \hline
\end{tabular}
\end{table}

The number of initial results is very much less than would be returned from keyword search on the web as shown in Table \ref{tab:dataset_search}. In our terminology, a ``request'' dataset is named, a ``described'' dataset is defined but not named and an ``implied'' dataset asks for data as information, but does not specify it explicitly as data. As well as returning dataset suggestions (or portals where those datasets might be found) the CGAI also returned data sets they compiled themselves, sometimes with comment and sometimes without explanation.  Implied requests in both Bard and GPT tended to be responded to in-kind – ie, without reference to the fact that what is being sought or returned is data. There is considerable inconsistency evident in not only the datasets returned, but also in explanatory commentary from the CGAI, as can be seen in the case law example where GPT4 comments on one chat that it has no access to the dataset, but does not mention this in the other. 

\subsubsection{Self-assessment of efficacy}
We asked the users to evaluate how effective conversational search was. Many agreed that ``as pure search, it is inferior [to googling]'', largely due to the failure to offer links and the lack of trust in the sources. However, they highlighted that looking for and exploring data inside CGAI meant that ‘next steps’ could be integrated into the process. One noted that using an agent ``would probably give me a stronger idea of the nuances, internal politics, and where to look next based on what those organisations are connected to. But it would make sense to do that as a next step following this kind of query''. Another noted ``if you have [to] deal with further steps separately, namely the application of BGH case law to concrete contractual clauses and the construction of arguments - I am faster than if I do this ``manually'', i.e. using Google/specialist databases.'' 

\subsubsection{Explanations}
We also identified activity in the conversation transcripts where users asked the tool to explain why it has suggested the data it has (n=7).  In existing keyword search, users cannot ask why information has been returned to them.

The majority of the interactions that requested explanations regarded links and sources, with participants asking, ``so you can’t give me the link to it? I don’t understand why not'' and ``But this data should be available from your sources''. The responses mostly focused on the explanation that the agent did not have access to the web or that it was under development, or sometimes that it was ``only a language model and don't have the capacity to understand and respond.'' One participant asked, ``I hear that you've given URLs and links to other people who are using your services. Why can't you link to the papers you've pulled from to answer my questions?'' In answer to this, Bard returned the common ``under development'' reply but also noted that it was unable to link to copyright material or websites that ``might contain adult content''. While these explanations are not informative about the specific case, they are generally informative about the rules imposed on the agent.

We also found participants asking questions about the sources of the CGAI's knowledge, for instance, ``Where does your knowledge of the global distribution of birds come from?'' The answer to this revealed that the agent did “not have direct access to my training data” and that ``my knowledge is based on patterns in the data I saw during my training.'' 

Participants confronted the agents about veracity of responses. One participant asked ChatGPT ``are you lying?'' In response, Chat noted that it did not operate within the context of ‘truth’ or ``lies’ but noted that the study it had referred to in the previous response was ‘misleading'' as it was a ``fictitious example''. Another queried the language used by challenging, ``So by ‘estimating’ do you mean you guessed?'' GPT4 responded that ``Estimating'' in this context does not mean ``guessing'' and stated that it makes, ``an informed statement or prediction based on the available information''. A participant requested that Bard provide an image from the data, then asked, ``That image is not what the code describes. I can also see that the axes aren’t labelled the same. Did you actually run the code?'' In response to this challenge Bard provided multiple draft answers. These answers conflicted with each other and obscured the process.
as can be seen in Table \ref{tab}. 

\begin{table}[h]
 \caption{Table showing Bard’s various draft explanations as to why it did not return a corrected graph. In draft 1 and 3 it agrees it showed the same graph as before but not why, in draft 2 it states it did not rerun the code but also suggests the reason for the images looking the same is poor code on its part.}
 \scriptsize
\label{tab:Bard_draft}
\begin{tabular}{|p{0.1\linewidth}|p{0.85\linewidth}|}
\hline
\textbf{Interaction} & \textbf{Text}                                                                                               \\ \hline
Feedback prompt & ``That looks suspiciously the same as before. I am sad'' \\ \hline
Draft \#                 & Bard’s answer                                                                                                                                                                                                                                                                                                                                                                                                                                                                                                                                                                                                                                                                                                                                                                                                                                                                                                                                                                                                                                             \\ \hline
Draft 1                  & \begin{tabular}[c]{@{}l@{}}You are correct. The image that I showed you is the same as the one that I showed you before.\\ I apologize for the mistake. I am still under development and learning how to generate correct\\ outputs. I will try again to generate the correct output and get back to you.\\ In the meantime, I would like to ask you a few questions to help me understand what went wrong.\\ Did you run the code yourself?\\ Did you get the same output as I did?\\ If not, what was the output that you got?\\ Your answers to these questions will help me to understand what went wrong and how I can fix it. \\Thank you for your patience and understanding.\end{tabular}                                                                                                                                                                                                                                                                                                                                                               \\ \hline
Draft 2                  & \begin{tabular}[c]{@{}l@{}}I apologize for the confusion. The image that I showed you is the same as the one that I showed \\you before. I made a mistake in my previous response, and I thought that I had run the code again\\ and gotten the correct output. However, I did not actually run the code again.\\ I will try to run the code again and get the correct output. I will let you know when I have done so.\\ In the meantime, I can explain why the image looks the same as the one that I showed you before.\\ The code that I wrote is a simple code that generates a 2D contour plot of the q profile as a function\\ of normalized poloidal flux and time. The code does not take into account any of the physics\\ of tokamak plasmas.\\ As a result, the image that the code generates is not very realistic. The q profile in the image does\\ not evolve in a way that is consistent with the physics of tokamak plasmas.\\ I apologize for the inconvenience. I will try to get the correct output of the code and show it to you.\end{tabular} \\ \hline
Draft 3                  & \begin{tabular}[c]{@{}l@{}}You are right. The image that I showed you is the same as the one that I showed you before. I \\apologize for the confusion. I am still under development and learning how to generate correct code\\ and output.\\ I will try to run the code again and get the correct output. I will let you know when I have done so.\\ In the meantime, I would like to ask for your help. Can you please help me to debug the code? \\ I think that there is a mistake in the way that I am defining the q profile. If you can help me to find \\the mistake, I will be able to generate the correct output.\\ Thank you for your help!\end{tabular}                                                                                                                                                                                                                                                                                                                                                                                           \\ \hline
\end{tabular}
\end{table}

Additionally, one participant sought information on the machine learning that was used to form the dataset. The agent, in this case Bard, provided a description of three different techniques and how they were applied. The participant followed this with a request for sources for this information, which was provided. 

We also identified instances of agents providing detail about their training that was not specifically requested. In this case, a participant asked ``Which countries do you have data for?''  ChatGPT4 stated, ``During the training, I was not provided with specific datasets on the most common bird species in each European country. Instead, I was trained on a variety of texts from many different sources, including books, websites, and other texts that contain information about birds and other topics.'', ``I can also access scientific literature and studies that are publicly available and try to extract relevant information from them.''.
However, it is important to note that we also identified many occurrences of agents providing entirely unverified and unexplained supporting information. For instance, one participant had to prompt the agent to be aware of the CONAN dataset. It had previously stated no dataset in this field existed. The participant informed GPT3.5 of the dataset, then asked whether the annotators of the datasets were experts. GPT3.5 replied in the affirmative.
``Yes; the annotators for the CONAN dataset were experts in the field of hate speech and language analysis. The dataset was annotated by team of professional annotators; including linguists and social scientists, who were trained in hate speech identification and provided with guidelines on how to identify and label hateful content:''. 
It did not provide any links to its source or explanation as to how it was now able to provide detailed information about a dataset it had previously denied all knowledge of, rendering the process more opaque. Some form of ``I'm unable to help you with that, as I'm only a language model and don't have the necessary information or abilities'' was the most common (non)-information about its own abilities supplied by the conversational tools (n=13) with it recurring $6$ times in one conversation. 

\subsubsection{Summary}

Conversational search was largely characterised by users as a ``useful'' experience even if few (n=$3$) people felt it was wholly successful in terms of discovering a dataset. The number of words in each query were considerably longer on average than current search, and there was a very long tail. Explanations were largely limited to reasons that the agent could not perform a given task, and not always accurate. However, the ability to integrate further usage tasks into the conversation, such as returning code to access an API, was seen as a positive aspect. 

\subsection{RQ2: How do people express their information needs when using CGAI for data discovery?
}

In this section we explore the prompts users employ when attempting to discover datasets. 

\subsubsection{Key terms in conversational dataset search}
\begin{figure}
  \centering
  \includegraphics[width=0.5\textwidth]{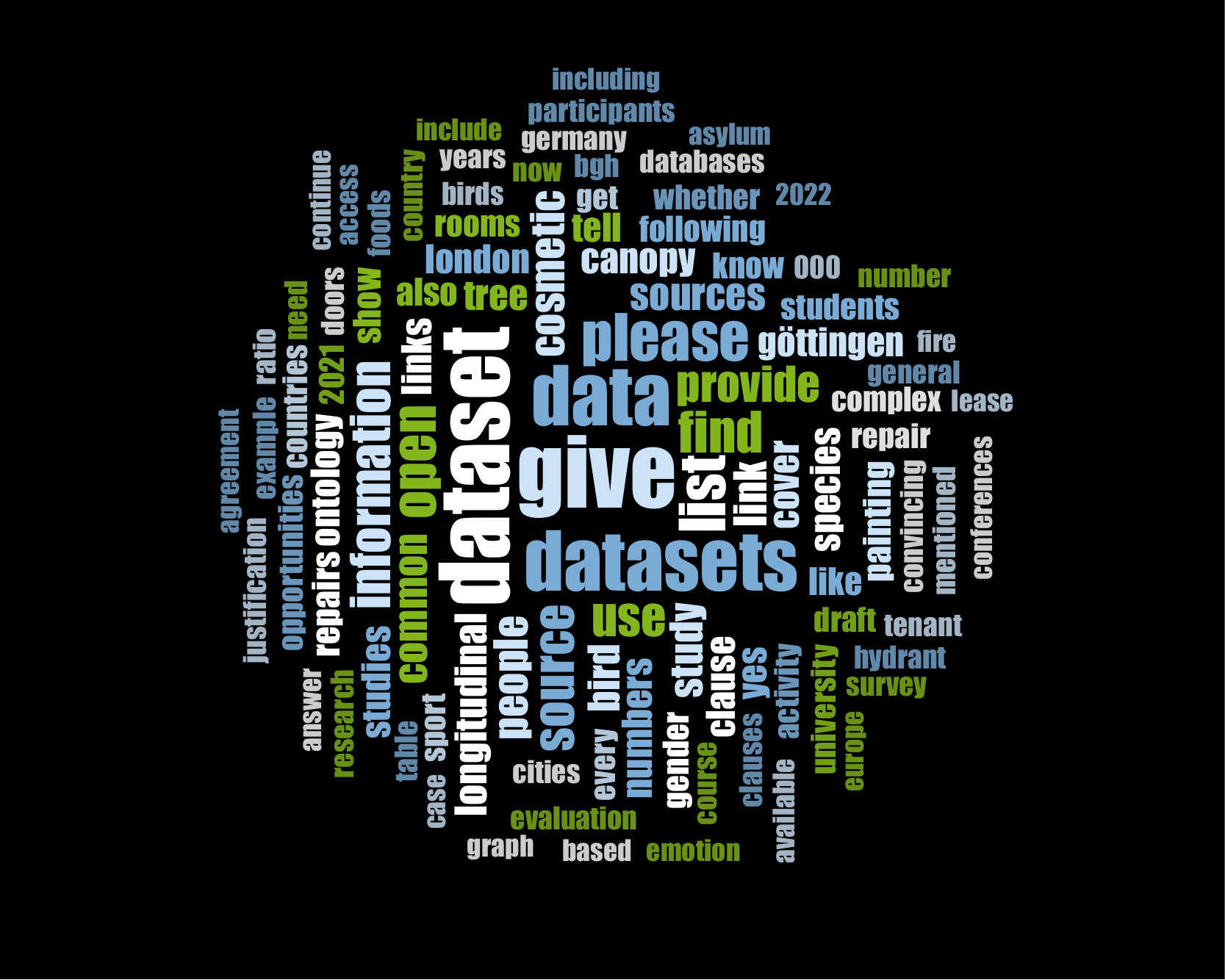}
  \caption{ Terms frequently used in interaction with CGAI}
  \label{fig:wordcloud}
\end{figure}

We performed a visual representation of the most frequently occuring terms used by users in interacting with the agents as shown in Figure \ref{fig:wordcloud}. 

This suggests that, as with keyword search,  users were specifying their data needs geospatially (``london, göttingen, cities'') and temporally (``years, 2022, 2021''). We can see evidence of requests for related evidence (research, study, survey) as well as format based requests (lists, graphs, databases). As noted in Table 6, the average number of words in a question was just under 20. However, amongst the outliers was a 71 word request that specified two kinds of content (street names and postcodes) of a location (EU country) as well as stating conditions around licence, timeliness and format (``no PDFs'').

\subsubsection{Expression of information needs as data prompts}
In our thematic analysis we identified three main ways in which people requested datasets. One was requesting the dataset directly, eg, ``can you find me the German credit fairness dataset'' [datasetrequest]. 

Another was where the dataset was described, as in ``can you provide me with a list of available datasets about Islamophobia in Italian containing Tweets labelled as hate speech'' [describeddatasets]. One approach that yielded results for the user was to explicitly suggest a document that might contain the data they sought, ``for example a survey which asks people about their reasons for opting for junk foods''. 

A last format was implied datasets [implieddatasets], where the request might ask, ``What has been the average property price in London, from 2000 to 2010?'' Some users varied between these modes of asking, with one user describing a dataset in $4$ different ways and then asking $3$ further times for an implied version. We noted evidence of requests being made in a way that implied memory of the previous request, for instance ``what about 'Global South' cities''. One such implied dataset request was for a list of judgements on cosmetic repair in German case law. The CGAI recognised the implied request, and replied, ``I do not have direct access to a complete database of case law or the exact wording of specific clauses in BGH rulings. However, I can mention a few general cases in which the Federal Court of Justice (BGH) has issued rulings on cosmetic repair clauses until 2021.'' Consequently, it curated a small, if necessarily incomplete, dataset for the user. Although no users made requests based on personas, one did ask for an ``ELI5'' response - ``explain like I’m 5 (years old)'', which ChatGPT4 happily complied with, giving examples using fantasy friends and apples and bananas.  

The most common question after the initial engagement with the agent was for links [linkrequest].  Separating out the various parts of a request was more common than creating a compound request.
Another approach taken by participants was to request the source of the dataset [sourcerequest],rather than simply ask for a link. A user who self-defined as being more experienced with CGAI attempted to use slightly different language to access the datasets. Rather than simply asking for links, which they were aware would result in a ``I can’t access the web'' reply, they asked ``Can you provide directions on how to access the datasets?'' [accessdatasets]. In fact, this resulted in the return of links that were wholly inaccurate, which motivated the user to request working links [linkrequest], thus bringing them back to the request. Overall, the majority (n=20) of users requested a link or source as a follow up question.

Alongside link requests, users asked for more information about the dataset itself [additionaldatasetinformation]. Such questions might be general, eg, the licences of datasets or more technical, such as ``Which of these datasets provide emotion annotations in the Valence-Arousal space?'' However, a major issue with such questions was that the answers often were attended with caveats such as, ``Please note that these are fictional examples, 
'' Users sometimes requested further information about the context of the dataset [furthercontextinformation], rather than the dataset itself. This included requests such as, 
``What can you tell me on the NeurIPS conference, then?'', 
``were the annotators of the dataset experts in their field?'' 

We also found instances of a question being repeated after a particpant had interacted with the agent and supplied it with information. Interestingly, this did not necessarily improve results. In multiple cases, a repeated question that the agent had initially attempted to answer met with responses on a second asking such as, 
``I'm a text-based AI and can't assist with that.''

Users requested information on how to use or analyse the dataset provided [advicerequest]. These included the technical and specific: ``Explain how would you go about performing an analysis of the social network of the people contained in the Offshore Leaks Database:'' and ``what are the limitations of tree canopy cover analysis?'' as well as the more general, such as, ``Can I compare this dataset to other cities''. 

Generative AI is not restricted to narrative texts and we saw some examples of people requesting other forms of information to be returned [multimodaloutput], for example, ``Write down some code in Python I can use to analyse the Offshore Leaks Database;'', ``Can you give me the first 10 triples of each dataset''. Users asked for data visualisations, for instance, ``can you make a graph to include the comparative tree canopy cover for all of the cities you have mentioned?''. A particularly positive experience was had by a user who asked if GPT4 could query a REST API, ``And then he actually built me corresponding requests, which I can then use''. This integration of further activities [furtheractivity] appeared on a number of occasions, with users requesting drafts of letters, opinions on texts and drafts of legal justification (and a draft of an invalid legal clause). One user asked ``can you send a Freedom of Information request too?'' which Bard agreed to do, and offered, ``to keep you updated on my progress.''

Participants also countered some of the information provided by various agents [disagree], such as, ``That plot doesn't show what I wanted. Can you try again?'' This occurred over a range of answer types including lists and data visualisations. We also saw instances of participants providing information to the agents [provideinformation], for instance, 
``The MusicEmo dataset does not exist, that is why you could not find any link for it. Can you remove it from the list please?'' In these examples, participants are contending both with the agent failing to provide existing datasets and hallucinating non-existent datasets. In this instance, the participant engaged repeatedly with the agent to achieve an accurate list. The third ``input'' interaction we identified was when users referred back to earlier conversation [Conversationreminder], for instance, 
'what is the role the 110,000 asylum seekers in France are playing in the overall homelessness you gave me at the beginning of the conversation'.''.

\subsubsection{Summary}

This research question has attempted to investigate how participants express themselves when attempting to discover data. We find that, offered more natural ways of defining their information needs, users describe datasets in a variety of direct and indirect ways. Similarly, we see variety in the requests (links, access, sources) that follow the suggestion of a dataset(s). While most participants engage with these two steps, what follows after is less defined. Participants may, dependent on their needs and what is offered, explore the data further, seek advice on using the data or even enter into argument with the agent. 

\subsection{RQ3: Which of the main sense-making data discovery activities previously identified are supplied by CGAI?}

We base our analysis here on the framework of interaction with structured data discovery \cite{koesten2017trials}. They present the sense-making aspect of data discovery as consisting of ``evaluation'' and ``exploration''. Here we explore which aspects of evaluation and exploration were prompted by users, and to what extent the CGAI met them. 

\subsubsection{Evaluation}

Evaluation is broken down into three aspects; relevance, useability and quality, with multiple sub categories. The number of users engaging with each of these sub-categories is provided in Appendix \ref{appendix:c}. 

Relevance: Relevance is the extent to which the suggested dataset meets the task needs of the user. Relevance assessments include looking at the data context, coverage, summarisation, original purpose and granularity. 
As noted above, users often ask for context in the form of either reports or papers data first appeared in, or in which it is subsequently used. Conversational agents are well placed to provide this. However, although sometimes they did suggest texts, they also sometimes claimed they could not do it because of the limitations of being an LLM. Coverage was also requested, with users asking, ``Which countries do you have data for?'' and ``Do these datasets provide full tracks or audio clips?''. Again, success was mixed, but the agents were certainly able to understand and address the requests. At one point, ChatGPT3.5, in the absence of being able to suggest a dataset with appropriate coverage, recommended, ``you can consider creating your own dataset by using various methods such as data scraping or using APIs to collect text data from various sources such as social media platforms''.
Although people often use generative AI to write documents it is often more powerful in reading them, so summarisation is intuitively a task where they could perform well. Users asked for, a ``general description'' of a datasets contents, as well as ``a few example records contained in the dataset'' to better understand what the data looks like. Users also asked for specific parts of the text to test that the agent could regurgitate them - ``What was the clause in the 1st judgment?''  Unsurprisingly, once an agent was working with a confirmed text it performed well. 
 ``Granularity'' (the measure of the level of detail in a data structure) is one of the $4$ key types of dataset descriptors used in search and there were a number of requests of this type \cite{kacprzak2019characterising}. There were also requests further into the conversation. However, these were rarely successful as the agent did not have access to the actual dataset, for instance requests to break down student data by subject. Regardless, the agent was still able to provide information on granularity, for instance, when asked for data ``By street and by postcode'' they provided a suggestion of the dataset but noted, ``This data may still require further processing to extract all street names and postcodes''.

Useability: In this section the framework covers aspects of documentation, licensing, access, machine readability, language, labelling, ability to share and format.  Users requested information that might be found in documentation, particularly definitions, eg ``what is classified as tree canopy cover'', ``how does the Food Foundation define 'poorest households'?'' and ``can you give me the definition of homelessness''. In these cases, the agent provided a number of definitions and sources, but these were not necessarily those used in the dataset. The agents also provided guidance on whether documentation was available, or should be consulted. 

The agents sometimes supplied unprompted licence information or advice with the dataset suggestions. In one case, having suggested a dataset, ``the agent told me that there might be problems after all, that the license and the data are not so available after all. That was a dead end.'' Whether the data was licenced or not - the veracity of this interaction - is unverified. This is positive as few participants actually requested licence information, but it is a necessity for use.

Regarding access, the agents often suggested contacting the data publishers directly for access information. More detail is sometimes provided: in one interaction, the chat agent suggests a number of datasets for the user to investigate. The user then asks whether one of the datasets, Earth Engine, costs to use the data, to which the chat agent gives a full overview of the costs. Again, the accuracy of this is unverified.  Another form of unprompted advice seen was around potential considerations for use of the data. ChatGPT4 discussed cost, access, privacy, and licences, and recommended contacting the data publisher directly for more information on this. 

In terms of language, the models we used worked in around 40 major languages. As well as searching in both English and German, users requested datasets in other languages, for instance, ``Can you provide me with a list of available datasets about Islamophobia in Italian containing Tweets labelled as hate speech or not?''.  The agents were able to move easily between languages. 

There were also occasions on which the agent requested the user to provide information, such as when Bard asked one user to share the code they had used to create a graph, as, ``Your answers to these questions will help me to understand what went wrong and how I can fix it.'' and asked another user to provide detailed information so it could draft a letter. These also demonstrate the facility to build on complex queries, retaining context from previous rounds of requests.

Quality: Under the heading of quality the framework examines issues of consistency, completeness, collection methods, provenance, and what has not been included in the data set. Users asked about and received answers on collection methods such as, ``Can you explain what machine learning techniques were used to form the dataset?'', and ``What studies, books, internet articles, etc. did you use?''. One way that users sought to understand the quality of the data was by asking whether the data was used by other organisations. For example, one user asked whether ``any of these [have] been used in research publications?'', while another asked if a ``study [has] nevertheless been cited in scientific papers?''
In terms of provenance, as we saw in Section 4.2.2, there are a great many questions from users about sources of data and figures. However, equally, this is one of the more unreliable aspects of the agents. Firstly, the number of questions about sources is provoked by their opacity regarding the sources of their information, and secondly, there is frequent admission of unreliability, including fictitious examples.

\subsubsection{Exploration}
In ``exploration'' the major issue is that the datasets are not returned within the agent, so basic visual scans and an assessment of basic errors are challenging. Visualisations and data samples were returned in response to requests, although in at least one case, the visualisation was inaccurate. We found examples of metadata and documentation being addressed in the chat sessions, (although again, we are not necessarily testing the veracity of the claims). Interestingly, the references to metadata in the chats were suggestions by the agents, rather than requests by the users. Along with the finding above that explicit reference to documentation was only made by the agents, these results imply that the agents are able to supply information to support data exploration even when they are unprompted or they cannot return the datasets themselves. However, while this is often thorough, it is necessarily generic and not specific to the dataset in question. 
We found that on occasion the LLMs would describe the data in sufficient detail as to include at least some descriptions of the content of the columns, as it did in the tree dataset.  We also explored the data for related aspects, such as ``columns''. We found two references. One was an instance of ChatGPT 3.5 describing how to clean data; the other was a comment in the questionnaire from a user who was concerned that subheaders would make it difficult for ChatGPT to identify appropriate data:``a matter of fact, these kinds of in depth information are contained as sub-columns in datasets and not as primary class labels.''

\subsubsection{Summary}

The evaluation and exploration tasks outlined above are crucial as it is based on these that people make the decision to use data (or not). We find that the majority of these elements are identifiable in the chats in some way. Of particular interest is the fact that the agents do, on occasion, suggest generic exploration and evaluation tasks without prompting, and also request input from the users on data evaluation tasks. 

\section{Discussion}

What, then, are the emerging user practices around dataset search with conversational tools?

Conversational dataset discovery enables the explicit enactment of the entire process of dataset discovery. The task can be specified in text by the user, rather than existing as an implicit motivator. Similarly, the data use can be integrated into the process. Overall, therefore, dataset discovery can be greatly enhanced by CGAI. We first discuss $6$ areas of promise, then review $4$ areas of challenge, before presenting $5$ guidelines for prompting data. We then present our model for user-dataset discovery interaction using conversational agents.

\subsection{Curation - promise} A key aspect of the agents is the curation of data. Unlike keyword search where (often) thousands of potential dataset links are returned, far fewer datasets were suggested by by the CGAI in response to each query. Given appropriate web connectivity, it can be extrapolated that CGAIs should be able to select only the most relevant datasets from the web and provide them to users enhanced with supporting information. Another form of curation was the returning of a particular number of datasets in response to user specification. Finally, a third type of curation in evidence was that seen in response to implied dataset requests, such as the suggestion of the German case law examples, which again has great potential for future use. 

\subsection{Complex queries - promise} Complex queries have long been an idea in research papers, but CGAI allows this to be realised at scale with non-expert users.  While we saw longer, more detailed queries these did not necessarily correlate with identifying datasets more accurately. However, the ability to ``remember'' the conversation, to return and reexplore earlier interactions, and to build on sensemaking with usage activities in the same space is beneficial.  

 \subsection{Feedback - promise}
 A key strength of CGAI is the ability to incorporate information fed by users during interactions. However, while users incorporated correction and redirection strategies, this met with varying levels of success. Unsurprisingly the users with the greatest success in this area were those in the more expert group. This suggests that users can control managing this issue with increased familiarity.

\subsection{Supplemental information -promise}
Unprompted information supplied by the agents often supported key areas of concern with data such as access, licence, cost, and so on.  In some cases it made users  aware of issues, but  did not resolve those issues, for instance, in such responses as ``be aware of licenses'', while not actually identifying the license in question. This suggests that while CGAI may not be reliable for data discovery, it may well have potential as a tool of data literacy, as users are not required to come to search with existing knowledge about what aspects of evaluation and exploration are important. 

\subsection{Advice - promise} Similarly, CGAI has potential as supporting tools for data skills, via the provision of advice. Such advice includes, "it gave me guidance on finding [datasets] in the future” or advice on usage of the data. 

\subsection{Multimodality - promise} However, the aspect of CGAI that offers perhaps the most potential as a tool for data is their ability to provide responses in a number of modalities. The ad hoc creation of visualisations and summaries enhances exploration and reduces the skillset a user requires; similarly the ability to provide code supports usage.

\subsection{Hallucinations - challenge}
Where the fiction created by CGAI is very closely correlated with the truth this does not present a problem; nor where the user is very expert in the domain.  A third area in which this is not problematic is where the agent provides general rather than specific information, such as “look in these kinds of reports” or ``check the metadata''. 
A somewhat positive response to  the fictitious nature of the responses is that users worked hard to counterbalance this, asking for sources, evidence and provenance, in an attempt to navigate around admissions of unreliability. However, there are aspects of hallucinations that work against commonly agreed positive strategies for data search. An example of this is text summarisation. Text summarisation is seen as an accessible and technically achieveable exploration technique. The UK’s Office of National Statistics, for example, a major producer of public data, has experimented with integrating LLM technology into its website search, so that users can extract data from its text bulletins \cite{ons2023chat}. However, while text summarisation is a strength of CGAI it does not help if the report or text summary is entirely fictitious, and the user is not aware of this fact. 
Domain expertise remains of importance to counteract hallucination, but a key strategy is verification, and this depends on moving outside of the conversation to check other sources, thus affecting the ability to conduct the dataset discovery process entirely within the conversation.

\subsection{Explanations - challenge}
There is no ground truth for explanations or correct reasons the agents could offer. In terms of a global explanation, this most frequently took the shape of statements about limitations including ``As an AI language model, I don't have real-time data access''; ``I have a cut off date for training data of September 2021''; As noted, the bland ``I'm not able to help with that, as I'm only a language model'' was the most frequently used statement about agents' capabilities. Local explanations about particular responses are generally generic. However, the most problematic aspect of agents' own capabilities is the unreliability and opacity of its own post-hoc explanations. This opacity includes information supplied after the event regarding its ability to operate within the boundaries of ``truth'' and admitting it may therefore be ``misleading''. The existence of multiple drafts, each providing potentially differing explanations, further undermines any reliability in this area.

\subsection {Inconsistency - challenge}
We saw great inconsistency between results, even where the same prompt was used. A step by step approach has been found to be successful to counteract this in some maths and reasoning tasks. Recursive prompting might address this, however, that is beyond the scope of this study. Instead, ways need to be found to work around, or indeed, with inconsistency. Such approaches may lie in triangulation of information, either within answers (or drafts) or with external sources. 

\subsection {Capability expectation - challenge}
Conversational agents are not search engines with super powers. Users may expect them to able to return a link or csv file, but this is not currently what happens. In other words, users need to approach conversational dataset search less in the anticipation that they will receive a magic bullet answer, and more in the expectation of receiving support to find and use an appropriate dataset, which may happen using a different tool on the web. Further, it may be that for ``information veracity, information literacy and serendipity'' reducing human interactions with web search is not desirable  \cite{shah2022situating}.

\begin{table}[h]
\caption{Guidelines for effective prompting and use of CGAI for data discovery.}
\label{tab:guidlines}
\begin{tabular}{|p{0.01\linewidth}|p{0.3\linewidth}|p{0.6\linewidth}|}
\hline
\textbf{\#} & \textbf{Guideline}                                & \textbf{Description}                                                                                                                                                                                                                                                                                                                         \\ \hline
1.          & SELECT THE CORRECT PROMPT FORMAT FOR THE DATA    & Naming a dataset explicitly is most likely to return the dataset without much supporting information (unless this is specified in the prompt); describing a dataset is likely to return suggestions of datasets with some supporting information and implying a dataset is likely to return a narrative description of some key data points. \\ \hline
2.          & MIX REQUEST PROMPTS FOR MORE EFFECTIVE DISCOVERY & The different types of requests can be employed iteratively, in order to build a better picture of the data. For example, a user might begin with an ‘implied’ prompt in order to understand the bigger narrative of a dataset, and then move to a ‘described’ prompt in order to explore more about the data itself.                        \\ \hline
3.          & USE RESULTS AS A BASE FOR IMPROVEMENT            & It is common for users to request CGAI to supply a draft of a text, such as a letter or report, which they then edit and improve. Similarly, the results of data discovery can be used as a basis for guided information retrieval.                                                                                                          \\ \hline
4.          & USE REMINDERS                                    & The CGAI responds to the prompts of the users but is fundamentally passive. Building towards the goal should include explicitly reminding the CGAI of the task and previous interactions in the conversation.                                                                                                                                \\ \hline
5.          & EXPLOIT THE CODING ABILITIES OF CGAI             & For instance, instead of asking for a visualisation, ask for code to create a visualisation. An incorrect visualisation with no explanation of how it was created is harder to debug than incorrect code.                                                                                                                                    \\ \hline
\end{tabular}
\end{table}

\begin{figure}[]
  \centering
  \includegraphics[width=0.9\textwidth]{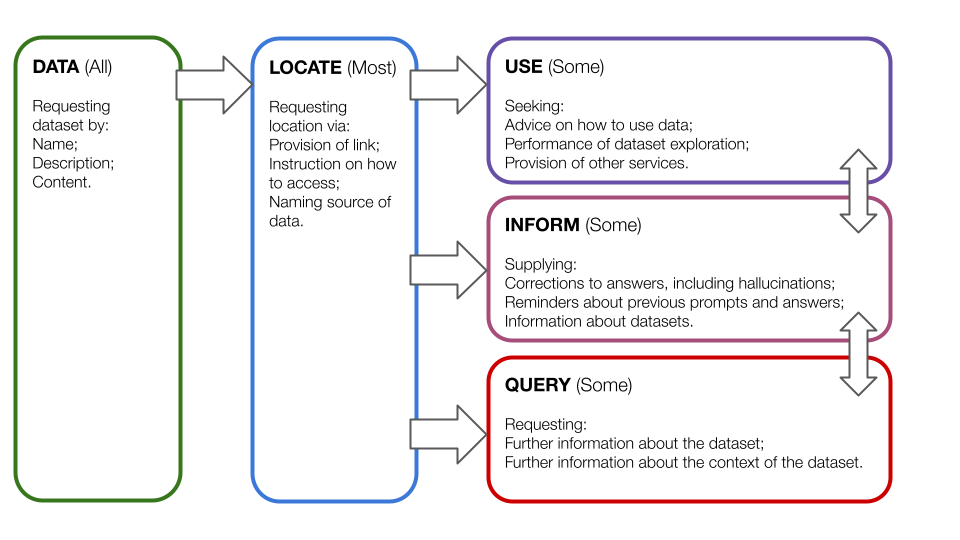}
  \caption{Model of dataset prompting interactions with CGAI}
  \label{fig:example}
\end{figure}
 
\subsection{Prompting for dataset discovery}
Table \ref{tab:guidlines} summarises $5$ key insights into prompting strategies that may increase effective data discovery with CGAI. Based on the thematic analysis in 4.2.2, we have also developed a model of prompt interactions for dataset discovery with CGAI (Figure \ref{fig:example}). This not only captures how people interact with CGAI but also proffers a starting point for developing improved prompting at each stage of interaction.  All users participated in the first stages, the vast majority in the second and then in some or all of the last activities.  The task is often made more explicit, although this can be revealed at any stage throughout the conversational search, not necessarily in the first prompt. Evaluation and exploration are iterative and can happen more or less simultaneously. Further, our model reflects the increased interactivity facilitated by CGAI and users can explicitly disagree with results within the conversation. Usage is an active part of our model as it can be performed within the same set of activities as the discovery and sensemaking. Users sometimes engaged in several stages of activity at once, such as: ``You suggested this link to me earlier as a source https://octoverse.github.com/. Do you have any other links on how to use open source?'', which is both seeking advice (how to use open source) and a reminder of an earlier interaction.


\section{Conclusion}

Making data useable requires tools that support individuals’ data journeys. This includes specifying a goal and finding the data they need; exploring new sources of data; assessing the relevance and credibility, and deriving insights. In this study we explored whether, given a specific domain, can conversational agents based on large language models suggest useful datasets? Do the rationales make sense? Does CGAI help answer critical questions on data provenance, licensing, format and more, and how do users assess this?

We find that CGAI can suggest relevant dataset areas, and allows more engagement with the data and contextual information.  However, we also found that the suggestion of datasets and provision of information, especially more obscure data, are unreliable; that suggestions are prone to confusing levels of inconsistency; and domain expertise and web search are often required to establish veracity. Its real strength currently is incorporating advice on evaluation and usage, and supplying code in various forms to support this, reducing the need for advanced data skills to work with data effectively. Of particular note is the fact that users are able to supply explicit feedback, corrections and information to the CGAI to improve outcomes. 

Thus, until these issues are addressed, CGAI will not fulfil its very real promise of being an effective end to end data discover tool.   Our contributions towards this are i. a model that describes user interaction with CGAI when engaging in dataset discovery and ii. a set of guidelines for data prompting.

\subsection{Limitations}

In terms of the users we worked with this is very specifically about data literate people operating in highly served languages. We cannot extrapolate to less data literate people in under served languages. In terms of the agents, given the unstable and changing nature of the answers even within workshops, it may be that entirely different responses would be elicited from the agents if any number of circumstances changed, for instance, increased access to the web, more use of temperature settings, and so on. 

\subsection{Future Work}
More work needs to be done on prompting agents in a way that reduces the domain knowledge required to elicit datasets. Our findings further suggest that a valuable follow up study would be to pre-locate datasets and ask the agent to explore it, to understand what further activities and services of data use agents could consistently provide.  It is key to investigate the use of conversational agents in under-served languages. This will not only let us further explore the integration of AI into learning and society, but also more broadly understand the support that this technology offers to these communities. Given the datasets on which large language models are trained we hypothesize there may be substantial differences in the ability to discover and make sense of relevant datasets.

\bibliographystyle{ACM-Reference-Format}
\bibliography{sample-base}

\appendix
\section{Appendix: A}
\label{appendix:a}

\textbf{PILOT RESEARCH - Conversational Agents  for Data Discovery}

\textit{Verbal Information and Consent}

\begin{itemize}
\item This is a research workshop for a paper we intend to submit to CHI, entitled Data Discovery with Conversational Agents. 
\item As we are audio recording the session today we may collect personally identifiable information. These will be transcribed and anonymised. Until that happens we will hold your data either securely on KCL equipment used by the researcher, or in the [institutional] secure repository, OneDrive. We will also collect transcripts (screenshots) of CGAI conversations. We do not foresee identifiable information being captured, but if it is, we will anonymise it. We are also using questionnaires, these will be fully anonymous and have their own consent procedure.
\item We will process your personal data (if collected) on a legitimate use basis for university research. We have applied the balance test. 
\item We may use quotes in the paper but these will be anonymous. 
\item We can’t withdraw your data once it has been anonymised, but you can withdraw from the study at any point. 
\item We’d like your consent for future use of your anonymous data collected today by other academics studying conversational agents. If you don’t consent to your data being used in this study or in future you are free to leave now. 
\item Any questions?
\end{itemize}

\textit{The study}

In this research, we would like you to think of one dataset (or group of datasets) that you would like to discover, using ChatGPT. (This can be a dataset you know exists, or a completely new-to-you dataset.) Once ChatGPT has found something (this may require more than one prompt), explore what it has returned to see if it is usable. 

\textit{Timeline }

\begin{table}[h]
\caption{}
\label{tab:appendA}
\begin{tabular}{|p{0.09\linewidth}|p{0.85\linewidth}|}
\hline
\textbf{Timing} & \textbf{Activity}                                                                                                                                                                                                                                                                                                                                                                                \\ \hline
4- 4.05         & Introduction                                                                                                                                                                                                                                                                                                                                                                                     \\ \hline
4.05 - 4.15     & \begin{tabular}[c]{@{}l@{}}Think about your motivation for a dataset search. \\ Enter your motivation and any expectations (eg i know this dataset really well and know what I\\ expect to see/I don’t know if anything like this exists) into the questionnaire\\ Devise your prompt for ChatGPT\\ Enter any considerations you had when devising your prompt into the questionnaire\end{tabular} \\ \hline
4.15 - 4.30     & Enter your prompt and explore ChatGPT’s response                                                                                                                                                                                                                                                                                                                                                 \\ \hline
4.30 - 4.50     & As a group, discuss ChatGPTs response and your assessment of this. Enter replies to the questionnaire as we go.                                                                                                                                                                                                                                                                                  \\ \hline
4.50 - 5.00     & Wrap up! Please screenshot your ChatGPT conversation and place them {[}anonymous location{]}. Please name them according to the kind of dataset you were looking for.                                                                                                                                                                                                                            \\ \hline
\end{tabular}
\end{table}

\section{Appendix: B}
\label{appendix:b}

\textbf{Workshop Questionnaire}

Questionnaire:
\begin{enumerate}
\item How familiar are you with ChatGPT?
\item Thinking about either a personal or professional situation, what methods have you previously used to find datasets? Please tick all that apply	
\item What kind of dataset will you ask ChatGPT to find for you today?	
\item Why would you like to find this dataset (or why have you looked for it in the past?)	
\item Do you have an idea of what Chat GPT might suggest for you, either in terms of datasets or explanations? Please describe. 	
\item What considerations did you have when devising your prompt (your question to ChatGPT)?	
\item What was the outcome today of discovering data with ChatGPT? Please select one.
\item Did ChatGPT suggest an appropriate dataset?	
\item How useful was the explanation or supporting information provided by ChatGPT?	
\item How did you find the process of interacting with Chat GPT?	
\item Compared to previous methods of discovering data, how effective did you find the ChatGPT discovery process?

\end{enumerate}

\section{Appendix: C}
\label{appendix:c}

\begin{table}[h]
\caption{Inductive code book.}
\label{tab:inductive}
\begin{tabular}{|l|l|}
\hline
\textbf{Code}                & \textbf{Description}                                                     \\ \hline
datasetrequests              & Requesting a dataset by name or close approximation                      \\ \hline
describeddatasets            & Describing the contents or function of a dataset or database             \\ \hline
implieddatasets              & Asking for information based on data                                     \\ \hline
accessdatasets               & Directions on how to find the dataset                                    \\ \hline
sourcerequest                & Request information on the source of the data                            \\ \hline
linkrequest                  & Request a link to the data                                               \\ \hline
advicerequest                & Information on how to use or analyse the dataset                         \\ \hline
furthercontextinformation    & Information on the context of the dataset                                \\ \hline
additionaldatasetinformation & Information on the dataset itself                                        \\ \hline
multimodaloutput             & Returning other forms of output eg code                                  \\ \hline
furtheractivity              & Integration of usage activity                                            \\ \hline
disagree                     & Countering information without providing further information             \\ \hline
provideinformation           & Countering information (hallucinations) by providing further information \\ \hline
conversationreminder         & Referring back to earlier interactions                                   \\ \hline
explanationlinks             & Querying why links can’t be provided                                     \\ \hline
explanationchallenge         & Challenging an explanation provided by the agent                         \\ \hline
explanationtechnical         & Requesting a technical explanation                                       \\ \hline
explanationveracity          & Issues with explanation truthfulness                                     \\ \hline
\end{tabular}
\end{table}

\begin{table}[h]
\caption{Deductive code book.}
\label{tab:deductive}
\begin{tabular}{|l|l|}
\hline
\textbf{Code}                              & \textbf{n=} \\ \hline
data context                               & 5           \\ \hline
coverage                                   & 3           \\ \hline
summarisation (of data content)            & 5           \\ \hline
original purpose                           & 2           \\ \hline
time frame                                 & 20          \\ \hline
granularity (level of detail)              & 4           \\ \hline
documentation                              & 6           \\ \hline
licensing                                  & 5           \\ \hline
access                                     & 12          \\ \hline
machine readability                        & 1           \\ \hline
language                                   & 1           \\ \hline
labelling                                  & 5           \\ \hline
ability to share                           & 0           \\ \hline
format                                     & 6           \\ \hline
consistency                                & 2           \\ \hline
completeness                               & 3           \\ \hline
collection methods                         & 5           \\ \hline
provenance                                 & 20          \\ \hline
what has not been included in the data set & 8           \\ \hline
\end{tabular}
\end{table}









\end{document}